\documentclass[notitlepage,twocolumn,fullpage,showpacs,superscriptaddress,showkeys,floatfix]{revtex4-1}
\usepackage[utf8]{inputenc}
\usepackage{amsmath}
\usepackage{amsfonts}
\usepackage{tabularx}
\usepackage{amssymb,bbold}
\usepackage{times,fullpage}
\usepackage{comment,float}
\usepackage{tcolorbox,framed}
\usepackage{array,graphicx}
\usepackage[colorlinks=true,citecolor=red,linkcolor=blue,urlcolor=blue]{hyperref}%
\usepackage{color,cancel}
\usepackage{csquotes}

\graphicspath{{./figures/}}

\begin{document}

\newcommand{\nwc}{\newcommand}
\nwc{\vs}{\vspace}
\nwc{\hs}{\hspace}
\nwc{\la}{\langle}
\nwc{\ra}{\rangle}
\nwc{\nn}{\nonumber}
\nwc{\Ra}{\Rightarrow}
\nwc{\wt}{\widetilde}
\nwc{\lw}{\linewidth}
\nwc{\ft}{\frametitle}
\nwc{\ben}{\begin{enumerate}}
\nwc{\een}{\end{enumerate}}
\nwc{\bit}{\begin{itemize}}
\nwc{\eit}{\end{itemize}}
\nwc{\dg}{\dagger}
\nwc{\mA}{\mathcal A}
\nwc{\mD}{\mathcal D}
\nwc{\mB}{\mathcal B}
\nwc{\col}[2]{\textcolor{#1}{#2}}

\nwc{\Tr}[1]{\underset{#1}{\mbox{Tr}}~}
\nwc{\D}[2]{\frac{d #1}{d #2}}
\nwc{\pd}[2]{\frac{\partial #1}{\partial #2}}
\nwc{\ppd}[2]{\frac{\partial^2 #1}{\partial #2^2}}
\nwc{\fd}[2]{\frac{\delta #1}{\delta #2}}
\nwc{\pr}[2]{$K(i_{#1},\alpha_{#1}|i_{#2},\alpha_{#2})$}
\nwc{\av}[1]{$\left< #1\right>$}
\nwc{\alert}[1]{\textcolor{red}{#1}}

\nwc{\zprl}[3]{Phys. Rev. Lett. ~{\bf #1},~#2~(#3)}
\nwc{\zpre}[3]{Phys. Rev. E ~{\bf #1},~#2~(#3)}
\nwc{\zpra}[3]{Phys. Rev. A ~{\bf #1},~#2~(#3)}
\nwc{\zjsm}[3]{J. Stat. Mech. ~{\bf #1},~#2~(#3)}
\nwc{\zepjb}[3]{Eur. Phys. J. B ~{\bf #1},~#2~(#3)}
\nwc{\zrmp}[3]{Rev. Mod. Phys. ~{\bf #1},~#2~(#3)}
\nwc{\zepl}[3]{Europhys. Lett. ~{\bf #1},~#2~(#3)}
\nwc{\zjsp}[3]{J. Stat. Phys. ~{\bf #1},~#2~(#3)}
\nwc{\zptps}[3]{Prog. Theor. Phys. Suppl. ~{\bf #1},~#2~(#3)}
\nwc{\zpt}[3]{Physics Today ~{\bf #1},~#2~(#3)}
\nwc{\zap}[3]{Adv. Phys. ~{\bf #1},~#2~(#3)}
\nwc{\zjpcm}[3]{J. Phys. Condens. Matter ~{\bf #1},~#2~(#3)}
\nwc{\zjpa}[3]{J. Phys. A ~{\bf #1},~#2~(#3)}
\nwc{\zpjp}[3]{Pramana J. Phys. ~{\bf #1},~#2~(#3)}

\title{Brownian gyration of an inertial ellipsoid}
\author{Soham Dutta}
\affiliation{University of Calcutta, 92 A.P.C. Road, Kolkata --- 700009, India}
\author{Arnab Saha}
\email{sahaarn@gmail.com}
\affiliation{University of Calcutta, 92 A.P.C. Road, Kolkata --- 700009, India}
\affiliation{Laboratoire de Physique Théorique et Modélisation, UMR 8089, CY Cergy Paris Université, 95302 Cergy-Pontoise, France}


\begin{abstract}

Recent studies on \emph{Brownian gyration} (BG) have focused primarily on spherically symmetric particles under overdamped conditions. To explore BG in the underdamped regime with a spherically asymmetric particle, we investigate the inertial dynamics of a microscopic ellipsoid in a dissipative medium. The particle is confined in a spherically-asymmetric trap and simultaneously coupled to two distinct thermal reservoirs. This configuration drives the system into a non-equilibrium steady-state (NESS) characterized by BG, which is quantified by the mean and fluctuation of the particle's specific angular momentum. Using inertial Langevin dynamics, we systematically analyze how this microscopic gyration depends not merely on the trap asymmetry and temperature difference, but also on the particle's intrinsic physical properties like shape and axial orientation, besides inertia. Our study uncovers fundamental differences between the gyration of spherical and non-spherical particles in overdamped as well as underdamped conditions, at microscopic scales. These findings provide key insights for optimizing Brownian gyration across a broader landscape of experimentally-tunable parameters.

\end{abstract}


\maketitle

\section{Introduction}

A spherical Brownian particle, confined in an anisotropic (spherically-asymmetric) trap, can be driven out of equilibrium when subjected to different temperatures along different degrees of freedom (DoF). As a result, the particle gyrates about the minimum of the trapping potential to produce steady-state currents. Asymmetry of the trap that couples the DoF and the finite temperature difference are indispensable features of the system to obtain a finite gyration current in the non-equilibrium steady-state (NESS). In recent times, a lot of research has been focused on such microscopic \enquote{machines} called Brownian gyrator (BG), which is able to produce directed motion or current under non-equilibrium conditions, even in the fluctuation-dominated microscopic regime.  

It has been shown in \cite{filliger2007brownian} that the BG can be mapped to that of a microscopic, steady-state heat engine, generating thermal torque. It has also been experimentally realized \cite{argun2017experimental}. Mancois et al. obtained the position probability distribution and the average angular velocity for the Brownian gyrator explicitly \cite{mancois2018two} in two dimensions (2D). A similar Brownian gyrator in a confining anisotropic parabolic potential has also been studied in \cite{dotsenko2013two}. The electrical analog of the BG was both experimentally and theoretically realized by \cite{chiang2017electrical}.  The power spectral densities of such electrical systems have been studied in \cite{cerasoli2022spectral}. In a similar context, the non-equilibrium thermodynamics of such systems (in particular, the heat flux and entropy production) have also been studied by \cite{ciliberto2013heat, ciliberto2017experiments}. A gyrating engine in the inertia-less regime with one rotational degree of freedom has been explicitly studied in \cite{siches2022inertialess}, where the emergent torque has an explicit dependence on the angle of rotation. The discussion on the autonomous Stirling engine \cite{izumida2018nonlinear, toyabe2020experimental} in \cite{siches2022inertialess} is also relevant in this context. The large-deviation properties and the associated statistics of entropy production in BG have been explored in \cite{mazzolo2023nonequilibrium}. The model for a BG as described in \cite{das2022inferring} consists of an anharmonic confining potential and the quantification of the entropy production rate has been done by considering the BG as a proto-typical microscopic heat engine. The work in \cite{dotsenko2022cooperative} has exhibited the co-operative dynamics existing between gyrating molecular tops. The case where the trapping potential of BG is non-harmonic has been explored in \cite{chang2021autonomous} and while it is time-dependent is studied in \cite{baldassarri2020engineered}. In \cite{squarcini2022fractional}, the BG has been subjected to non-trivial fractional Gaussian noises. The concept of an effective temperature is also introduced to BG in \cite{cerasoli2018asymmetry}. Enhancement of the efficiency of shear-driven gyrators has been discussed in \cite{abdoli2025enhanced}. However, while the physics of overdamped, spherical Brownian gyrators is now well-studied, real-world microscopic systems often possess inherent inertia and geometric complexities. Moving beyond the spherical archetype, here we explore how shape anisotropy and inertial dynamics together alter the landscape of microscopic gyration.

Theory of BG can be further developed once we consider the spherically-asymmetric colloidal particles. In general, the stochastic thermodynamics of a spherically-asymmetric particle crucially depends on its geometry and orientation \cite{dutta2026stochastic}. In case of BG also, the dynamics as well as thermodynamics of the gyrating particle should depend on them. Recently, it has been shown that in case of a spherically-asymmetric, overdamped BG, the confinement itself does not need to be spherically-asymmetric. Instead, the DoF of the particle are coupled by the inherent properties --- such as its orientational bias and shape anisotropy. Hence, in this case, one can dispense the spherical asymmetry of the external trap to obtain microscopic gyration. However, different DoF of the gyrating particle still have to be in contact with the baths at different temperatures to drive the system out of equilibrium \cite{dutta2026microscopic}.    

Although microscopic gyration is being explored extensively in viscosity-dominated, overdamped colloidal systems, the theoretical framework can be further developed to more realistic systems where the particle has considerable inertia. However, fewer studies have been conducted so far in the inertial regime. The underdamped Langevin dynamics of a microscopic gyrator is studied in \cite{mancois2018two}, besides the overdamped one. The stochastic energetics of an inertial gyrator is discussed in \cite{bae2021inertial}. A mesoscopic magneto-heat pump made of a single charged particle is analyzed in \cite{abdoli2022tunable}. Memory-induced, magnetic behavior of an active, visco-elastic gyrator is studied in \cite{muhsin2025active}. Expanding the research on BG with inertial stochastic dynamics should open up the possibility of finding microscopic gyration even in dusty plasma \cite{wang2018structures}.


The systems relevant to the current work are the spherically-asymmetric, confined microscopic particles with considerable inertia. The particle can act as a minimal stochastic machine, generating a systematic gyration current when driven out of equilibrium, under the influence of thermal noises from different heat baths simultaneously. Here, we will explore how the microscopic gyration depends on the inherent properties of the particle, such as, its spherical asymmetry, axial orientation and inertia. The particle that we consider here, experiences asymmetric, tensorial frictional drags from the surrounding medium due to its bi-axial shape \cite{dhont1996introduction,berg1993random,kim2013microhydrodynamics}. It is this tensorial character of friction for which the translational DoF of the particle become inherently coupled with each other in the (fixed) laboratory frame \cite{dhont1996introduction,han2006brownian,ghosh2020persistence,dutta2026microscopic,dutta2026stochastic}, along with an inevitable coupling between translational and rotational DoF, assisted by an orientational bias or tilt of the anisotropic particle at a finite angle. It will be illustrated here with a spherically-asymmetric, bi-axial microscopic particle that gyrates in a noisy environment, resulting in a NESS which depends explicitly on this dissipative coupling. 


The paper aims at studying the kinematics of an inertial, anisotropic gyrator at the micro-scale to exhibit how it is regulated by the shape, orientation and inertia of the particle --- all being the intrinsic properties of the particle itself. The results show prominent qualitative as well as quantitative differences from their isotropic, overdamped counterparts, due to the interplay of its shape and orientation-dependent, dissipative coupling together with its inertia. The simultaneous introduction of shape, orientation and inertia in the dynamics of a gyrator opens multiple avenues to optimize the gyration. It will also be shown here that such an anisotropic gyrator, when electrically charged and subjected to an external magnetic field, can even become a \enquote{magneto-gyrator} \cite{abdoli2022tunable,muhsin2025active} and the resultant orbital magnetic moment will eventually depend on its geometry and inertia. Besides extensive numerical quantification, in the regime of small shape anisotropy, the model will also be treated analytically.    



The paper is organized in the following manner. We consider a microscopic ellipsoid with finite inertia, trapped in an asymmetric potential, and subjected to unequal temperatures along the two translational DoF. After discussing the Langevin dynamics corresponding to our model, we numerically simulate the anisotropic system in order to study its steady-state gyration via the average specific angular momentum. After studying the variation of the kinematics of gyration with the relevant system parameters, we studied the performance of the gyrator by numerically quantifying its mean-to-fluctuation ratio and steady-state distribution of specific angular momentum. Several ways have been proposed in order to reduce the effects of fluctuations on the output. The heat fluxes between the orthogonal DoF of the gyrating ellipsoid has also been studied briefly. The subsequent section then deals with the analytical results obtained in the regime of small shape anisotropy. The equations of motion are linearized using relevant approximations and the steady-state covariance matrix is obtained therein. The analytical result for specific angular momentum in this regime agrees well with the numerical results of the anisotropic system, when the latter is subjected to the stipulated range of parameters decided by the approximations. In the last section, by introducing a magnetic field, the gyration-induced magnetic moment and its dependence on the shape of the ellipsoid is studied numerically, to identify the magnetic and non-magnetic regimes. Finally, the paper concludes with the summary of the results obtained.

\section{Confined inertial ellipsoid --- The model}

\begin{figure}[htp]
    \centering
    \includegraphics[width=8cm]{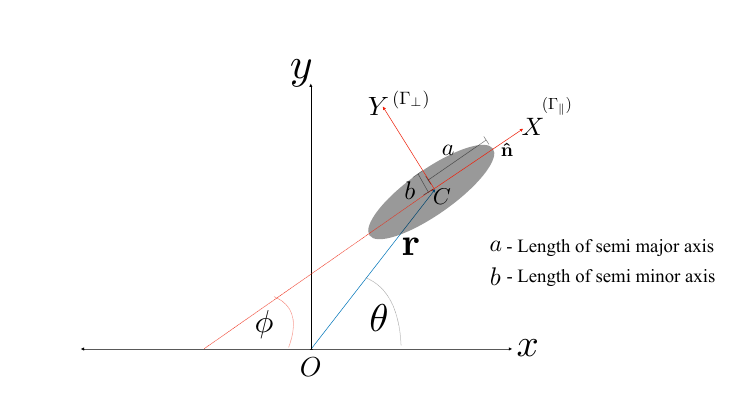}
    \caption{Schematic diagram of the various co-ordinates associated with an ellipsoid in the $xy$-plane. The body frame is denoted by  $(X,Y)$ and the lab frame is denoted by $(x,y)$. Here, ${\bf{r}}=\{x,y\}$ is the position vector of the center-of-mass (C) of the ellipsoid [with respect to (w.r.t.) the lab frame] and $\theta$ is the corresponding polar angle. The angle $\phi$ denotes the axial orientation of the ellipsoid (w.r.t. the lab frame). This is also the angle between body and lab axes. The major axis of the ellipsoid is along $\hat {\bf n}$, i.e. the unit vector along $X$. The axial mobilities/inverse frictional drags $\left(\Gamma_\parallel \equiv \frac{1}{\gamma_\parallel},\Gamma_\perp \equiv \frac{1}{\gamma_\perp}\right)$ are marked along the respective body-axes.} 
\label{ellipsoid}    
\end{figure}

The model consists of an ellipsoidal particle moving with considerable inertia. 
The dynamics of the particle is restricted to the $xy$-plane (see Fig.[\ref{ellipsoid}]). Its center-of-mass (denoted by 'C' in Fig.[\ref{ellipsoid}]) is translating across the plane, being confined in an optical trapping potential. Translational dynamics of the center-of-mass involves all the forces acting on the ellipsoid. 

Apart from the translation, the particle can independently rotate with the axis being perpendicular to the $xy$-plane and passing through C. This rotation is denoted by the angular orientation $\phi$, relative to the $x-$axis of the lab frame (see Fig.[\ref{ellipsoid}]). As the particle is spherically-asymmetric and bi-axial, the axial orientation can also be denoted by the unit vector $\hat {\bf n}=(\cos\phi, \sin\phi)$, which is along its major axis. The rotation or the dynamics of the orientation of the ellipsoid is such that all the torques acting on the ellipsoid are balanced \cite{simpson2007optical,la2004optical}. This aspect will be revisited in greater detail, later in this section.


The equations of motion that completely specify the inertial Langevin dynamics of the ellipsoid (of mass $m$) in the lab frame are,

\begin{equation}
m(\partial_t v_i)=-\gamma_{ij}v_j-\partial_i U + \xi_i(t)
\label{eom1}
\end{equation}

\begin{equation}
\partial_t \phi=-\Gamma_r k_\phi (\phi-\phi_0)+\xi_r(t)
\label{eom2}
\end{equation}

where, $v_i=\partial_tx_i$, with $i\in \{x,y\}$, $j\in \{x,y\}$, and $\phi \in [0:2\pi]$. For details on the derivation of translational equations of motion, one may consider Appendix-I.  The orientational dynamics is considered to be overdamped \cite{han2006brownian,chaki2024dynamics,dutta2026microscopic,dutta2026stochastic}, whereas, the translational dynamics is underdamped \cite{montana2023inertial}. In this limit, due to the small size of the particle, the moment of inertia of the particle is taken to be negligibly small, and therefore, the angular momentum relaxes faster than the linear momentum \cite{lowen2020inertial}. It may be noted here that the  $m\rightarrow0$ limit of Eq.[\ref{eom1}] yields the overdamped Brownian dynamics of a colloidal ellipsoid \cite{han2006brownian}.

\subsection{Translational dynamics}

According to Eq.[\ref{eom1}], the particle is subjected to three distinct forces during its translation in two dimensions (2D) ---

(i) The viscous drag, $-\gamma_{ij}v_j$, where $\gamma_{ij}$ is the friction tensor. While navigating through a fluid, an ellipsoid experiences anisotropic friction, owing to its bi-axial shape. On a plane, hence, $\gamma_{ij}$ will be a $2\times 2$ symmetric matrix, the components of which are explicitly dependent on the shape and orientation of the particle as \cite{han2006brownian,dhont1996introduction,chaki2024dynamics,montana2023inertial}:

\begin{eqnarray}
\nonumber
\gamma_{xx}&=&\gamma_{\parallel}\cos^2\phi+\gamma_{\perp}\sin^2\phi\\
\nonumber
\gamma_{yy}&=&\gamma_{\parallel}\sin^2\phi+\gamma_{\perp}\cos^2\phi\\
\gamma_{xy}&=&\gamma_{yx}=\Delta\gamma \sin\phi\cos\phi
\label{gamm}
\end{eqnarray} 

where, $\gamma_{\parallel}$ and $\gamma_{\perp}$ are the longitudinal and transverse frictional drags \cite{berg1993random} of the ellipsoid, along the directions parallel and perpendicular to its major axis respectively. Here, $\gamma_{\perp}>\gamma_{\parallel}$, and they depend on the ratio of the lengths of major and minor axes of the ellipsoid \cite{dhont1996introduction,berg1993random}. The frictional drag difference, $\Delta\gamma\equiv\gamma_{\parallel}-\gamma_{\perp}$, becomes negative as a consequence. 

If the particle is spherically-symmetric, then $\gamma_{ij}$ becomes diagonal with $\gamma_{\perp}=\gamma_{\parallel}$. Hence, $\Delta\gamma$ is a measure of the geometric asymmetry of the particle.  The off-diagonal terms containing $\Delta \gamma$ couple $v_x$ and $v_y$ in the equations of motion, besides the translation-rotation coupling, which is an inevitable feature of bi-axial particles \cite{dhont1996introduction}. However, the Langevin equation for $\phi$ has no dependence on its translational counterparts whatsoever.

\begin{figure*}[htp]
    \centering
    \includegraphics[width=6cm]{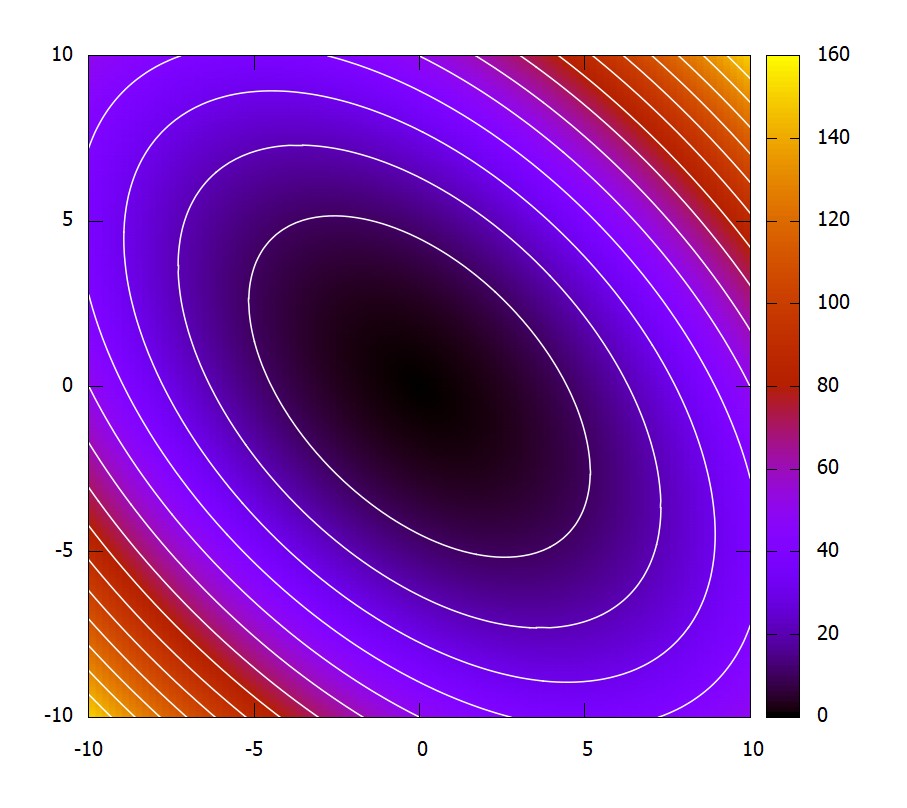}
    \includegraphics[width=10cm]{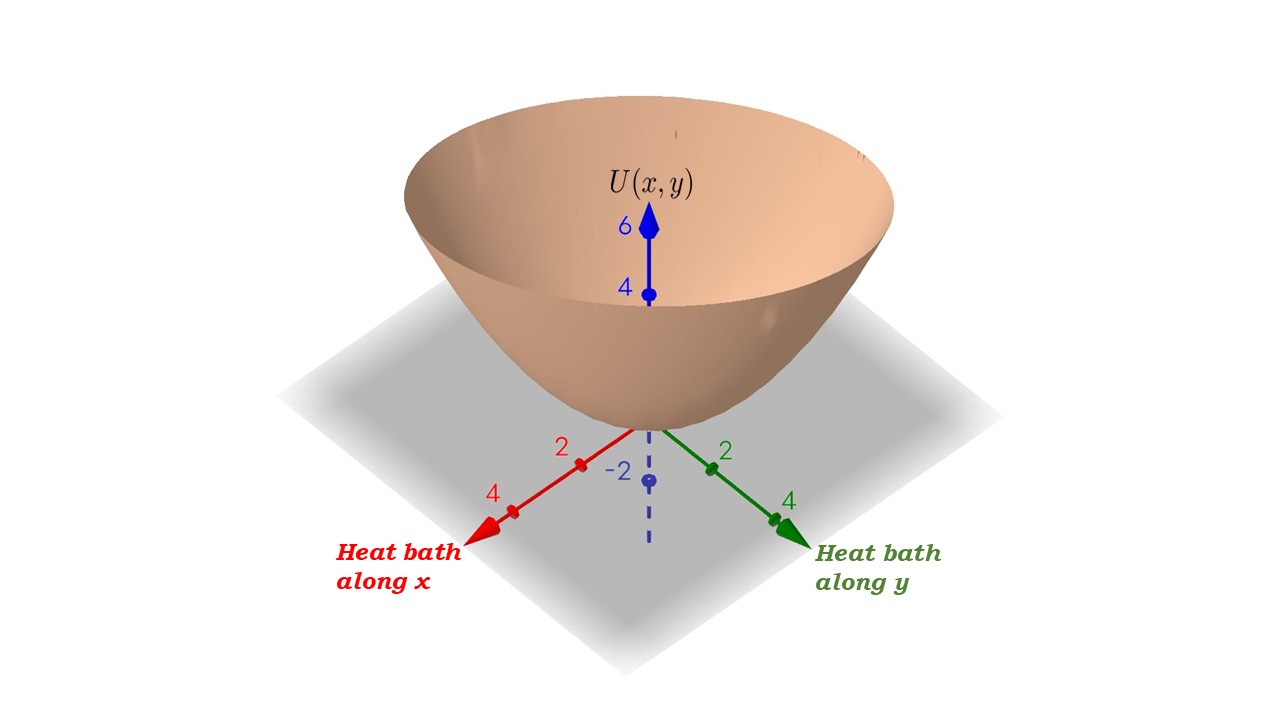}
    \caption{(Left) 2D contours of the potential $U(x,y)$, for $k=1$ and $\alpha=0.5$. (Right) Projection of $U(x,y)$ in space, for $k=1$ and $\alpha=0.5$. Note the mismatch of potential and temperature axes --- the potential landscape is not \enquote{aligned} with the heat baths. This landscape becomes right-aligned when the sign of $\alpha$ is flipped, thus acting as the \enquote{roadmap} of gyration.}
\label{trap}    
\end{figure*}

(ii) The trapping force, originating from an asymmetric spatial confinement \cite{bae2021inertial,mancois2018two,muhsin2025active,dotsenko2013two}, $U(x,y)=\frac{k}{2}\left(x^2+y^2\right)+\alpha xy$ (where, $k$ and $\alpha$ are the stiffness constant and asymmetry parameter respectively, with $|\alpha|<k$ required for stability), which spatially bounds the ellipsoid and also couples its center-of-mass coordinates ($x,y$). Due to the asymmetry ($\alpha\neq0$)  of the potential, the principal axes of the potential are inclined along the left or right diagonal depending on whether $\alpha$ is positive or negative (see Fig.[\ref{trap}]). The sign of $\alpha$ and the difference of temperatures along $x$ and $y$ together determine the direction (clockwise or anti-clockwise) of gyration. The asymmetry parameter $\alpha$ and the temperature difference both are essential ingredients for gyration \cite{muhsin2025active,bae2021inertial,mancois2018two}. The heat flux generated during gyration due to the temperature difference drives the system towards a NESS. The origin of the two orthogonal heat baths (being simultaneously connected to the particle) comes from thermal noises, as discussed below.


(iii) Fluctuating forces $\xi_x(t)$ and $\xi_y(t)$ are the zero-mean Gaussian, white, random noise components along $x$ and $y$ respectively. The temperatures along $x$ and $y$ are $T_x$ and $T_y$ respectively, with $T_x\neq T_y$. The gyration is a direct evidence of a NESS established due to two distinct thermal reservoirs and the heat flux facilitated between them. The thermal noises are such that \cite{han2006brownian,chaki2024dynamics,mancois2018two,muhsin2025active,abdoli2025enhanced} : $\langle\xi_i (t) \rangle=0$, $\langle\xi_i (t) \xi_j(t')\rangle=2\sqrt{T_i T_j} \gamma_{ij}(\phi) \delta(t-t')$. Note that the noises along the translational DoF are not independent entities now, but are correlated (in the lab frame) due to $\gamma_{xy}$. Here, $\boldsymbol{\xi}(t)=\sqrt{2}\bar{\bar{D}}(\phi)\boldsymbol{\eta}(t)$ \cite{chaki2024dynamics,han2006brownian,ghosh2020persistence}, where, $\boldsymbol{\xi}(t)=\{\xi_x(t),\xi_y(t)\}$ and $\boldsymbol{\eta}(t)=\{\eta_X(t),\eta_Y(t)\}$ are the translational noises in the lab and body frames respectively. In our model, the quantity $\bar{\bar{D}}(\phi)$ is a $2\times2$, orientation-dependent \enquote{diffusivity} tensor, given as :

\begin{eqnarray}
   D_{ij}\equiv\begin{bmatrix} \sqrt{\gamma_\parallel T_x}\cos\phi & \sqrt{\gamma_\perp T_x}(-\sin\phi)\\\sqrt{\gamma_\parallel T_y}\sin\phi & \sqrt{\gamma_\perp T_y}\cos\phi \end{bmatrix}
   \label{diff}
\end{eqnarray}

This ensures that the translational DoF are equilibrated to the respective bath temperatures at large times, in the absence of  the asymmetry of the potential that couples $x$ and $y$. Here, $\langle\eta_p (t) \rangle=0$ and $\langle\eta_p (t) \eta_q(t')\rangle=\delta_{pq} \delta(t-t')$, with $p\in\{X,Y\}$ and $q\in\{X,Y\}$.


We emphasize here that the particle \enquote{feels} distinct temperatures (via different thermostats \cite{abdoli2020correlations}) along the different DoF. In experiments, there can be several ways to implement this. Fluctuating electrical voltage and electrical heat bath are theoretically proposed in \cite{filliger2007brownian}. It is experimentally implemented in several recent studies (e.g. \cite{martinez2013effective,martinez2017colloidal}). As proposed in \cite{filliger2007brownian}, the advantage of an electrical signal to mimic the temperature is that, it can be applied along a particular direction and thereby along a particular DoF of the particle. Such experimental realizations suggest that these temperatures are \enquote{effective} temperatures, since in experiments they may not be originated from the thermal energy of the ambient fluid in which the particle resides, but arises from an effective stochastic force that can be applied through a variety of externally-modulated mechanisms.

\subsection{Orientational dynamics}

In our model, the dynamics of axial orientation $(\phi)$ of the ellipsoid (as described in Eq.[\ref{eom2}]) is governed by the sum of two torques --- a deterministic one and a thermally-generated, stochastic one. The former is a restoring torque \cite{dutta2026microscopic,dutta2026stochastic}, i.e. $\tau(\phi)=-k_{\phi}(\phi-\phi_0)$, which is linear in $\phi$ and tries to restore the orientation of the particle to $\phi_0$, with $k_\phi$ being the torque-strength. 
The latter is a stochastic torque which can be originated from thermal fluctuations. It is modeled by delta-correlated, Gaussian, white noise with zero mean, such that: $\langle\xi_r (t) \rangle=0$ and $\langle\xi_r (t) \xi_r(t')\rangle=2\Gamma_r T_r \delta(t-t')$. Here, $\Gamma_r\equiv\frac{1}{\gamma_r}$ is the (scalar) mobility/inverse friction coefficient and $T_r$ is the temperature (which can, in general, be different from $T_x$ and $T_y$) corresponding to the orientational dynamics. The rotational mobility is introduced due to the viscosity of the surrounding fluid in which the particle rotates. As the dynamics of $\phi$ is governed by a linear equation, its steady-state distribution will also be Gaussian. The first two moments of $\phi$ --- namely, mean $(\langle\phi\rangle)$ and variance $(\sigma_\phi^2=\langle\phi^2\rangle-\langle\phi\rangle^2 )$, are then sufficient to completely specify the distribution. At large times, the exponential transients are decayed and these moments acquire the following forms: $\langle\phi\rangle=\phi_0$, $\sigma_\phi^2 = \frac{T_r}{k_\phi}$. It is evident that a large value of the restoring torque-strength suppresses the fluctuations in $\phi$, for a given $T_r$. For a non-zero value of $\phi_0$, the Gaussian distribution becomes asymmetric about $\phi=0$. This will act as an aid in the survival of velocity-coupling in the translational dynamics, averaged over all possible $\phi$. This will be evident in the subsequent sections.

At the scale of a single colloidal particle, forces and torques both can be applied by irradiating it with laser via {\it{optical tweezers}} \cite{ashkin1970acceleration,ashkin1986observation,ashkin1987optical,lehmuskero2015laser,bowman2013optical,simpson2007optical,deufel2007nanofabricated,ling2010optical,simpson2011computational,callegari2015computational,loudet2014optically,barton1989theoretical,borghese2008radiation,roy2016using,hoang2016torsional,bang2020five}. Since their introduction, optical tweezers are extensively used to confine and manipulate micro- and nano-sized, living as well as non-living objects, and thus, to study their fundamental properties. Various theoretical as well as experimental methods are developed to estimate the forces and torques produced by the tweezers. Different forces such as, intensity-gradient force \cite{ashkin1986observation} and phase-gradient force \cite{roichman2008optical}, scattering force and radiation pressure \cite{ren1994radiation} are involved in an optical tweezer. They are the different aspects of changes in electro-magnetic momentum due to light-matter interaction. The interplay between the electromagnetic stress (quantified by the Maxwell stress tensor) of the laser beam used in the tweezer and the particle trapped inside the beam via these forces can be tuned in such a way that the particle feels an effective, linear restoring force towards the focus of the beam where the intensity attains a maximum value \cite{simpson2007optical,bowman2013optical,liang2020simultaneous}. This restoring force creates a spatial confinement for the particle. Importantly, the forces obtained from the optical trap can even generate torques influencing the orientational dynamics of the confined particle in a fluid or even in vacuum \cite{bruce2021initiating,millen2020optomechanics}. Polarization of the incident beam combined with the optical properties of the confined particle can be the crucial factors to determine the features of the torque. For example, linearly polarized Gaussian laser beam can exert conservative, restoring torque on a particle made of typical uniaxially-birefringent materials, and therefore, the particle can be \emph{angularly-trapped} in a harmonic potential in the $\phi-$space \cite{la2004optical,deufel2007nanofabricated,hoang2016torsional,bang2020five}. Another way to control the orientation of a trapped, non-spherical particle is by using the laser with an asymmetric beam profile. It also exerts an additional conservative torque on the particle, originated from the asymmetric gradient force, irrespective of the beam polarization. A common asymmetric beam is the elliptic Gaussian beam, which will produce a gradient torque to restore the orientation of the particle towards the long axis of the elliptical laser beam \cite{o2002rotational}. These experimental instances form the major motivation in introducing $\tau(\phi)$ in Eq.[\ref{eom2}], and then to approximate it as a linear, restoring torque.

\section{Numerical results for the fully anisotropic system}

\subsection{Generic scheme of simulation}

The Langevin equations for translation and orientation, as given in Eq.[\ref{eom1},\ref{eom2}], have been simultaneously discretized and numerically integrated to obtain velocities and positions using the finite time-difference scheme, with a time step-size of $\Delta t=10^{-3}$. The total number of iterations are $10^{7}$, out of which $10^{6}$ time-steps are taken as the transient regime.  The required time-averaging has been done beyond this interval. Gaussian random numbers have been generated with zero mean and unit variance for the random noises. The orientation angle $(\phi)$ has been subjected to periodic boundary condition between $[0:2\pi]$. The following values of the system parameters are considered wherever they are required to be constants, unless mentioned otherwise: $\gamma_r=1, |\Delta\gamma|=9, T_x=1, T_y=10, T_r=1, k=1, \alpha=0.5, k_{\phi}=50, \phi_0=\frac{3\pi}{4}, m=1$. Throughout the simulation, the values of $T_r$ and $\Gamma_r$ have been kept equal to the lowest among $\{T_x,T_y\}$ and $\{\gamma_\parallel,\gamma_\perp\}$ \cite{mandal2024diffusion} respectively.

\subsection{Dynamical cross correlation: coupling of translational DoF}

Before going into the kinematics of gyration, it is important to explore the dynamical couplings in Eq.[\ref{eom1}] --- namely, coupling of the velocity components by friction tensor and coupling of the position coordinates by the asymmetry parameter $(\alpha)$ of the trap, since it is a combined effect of the couplings between the DoF and the difference of temperatures along them. The effects of translational DoF on one another must be thoroughly probed to understand the gyratory response. A quantity which is relevant for this purpose is the steady-state cross-correlation, $\langle xy \rangle$. This quantity vanishes for a spherically-symmetric external trap in case of a point or a spherically-symmetric Brownian particle, signifying the vanishing microscopic gyration of the same \cite{mancois2018two,bae2021inertial,dotsenko2013two}.

\begin{figure}[htp]
    \centering
    \includegraphics[width=8cm]{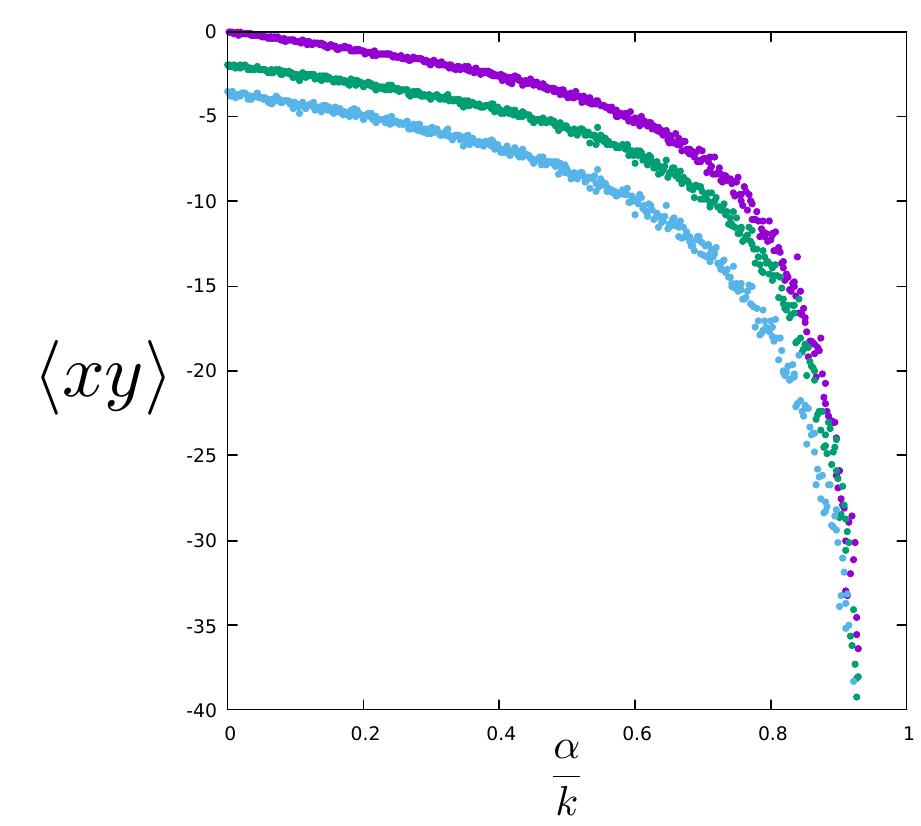}
    \caption{Plot of $\langle xy \rangle$ versus $\frac{\alpha}{k}$ (with $\alpha>0$), for various $|\Delta\gamma|$. The purple, green and blue points are respectively for $|\Delta\gamma|=0,4,9$.} 
\label{xyavg}    
\end{figure}

However, in our model, this can be counteracted by the anisotropic shape --- Fig.[\ref{xyavg}] shows that the correlation vanishes only when $\alpha=0=|\Delta\gamma|$, i.e. for a spherical particle in a purely harmonic trap. For $\alpha=0$ and $|\Delta\gamma|\neq0$, the magnitude of this correlation increases in proportion with the shape anisotropy. As $\alpha\longrightarrow k$, the correlation sharply diverges. This divergence will have a detrimental effect on the gyration, which will be detailed in the subsequent sections. However, we mention here that to obtain a stable gyration, $|\alpha|<k$ must be strictly maintained. In the midst of a finite temperature difference, how the sole presence of $\alpha$ drives gyration has been studied in \cite{muhsin2025active,bae2021inertial,filliger2007brownian,mancois2018two,dotsenko2013two} --- the trap asymmetry continually drives the particle between two heat baths, generating a non-equilibrium gyration current. The next sub-section will analyze how the inertia of the particle together with an additional coupling (introduced via the shape asymmetry of the particle) largely affects this motion.

\subsection{Kinematics of gyration --- specific angular momentum}

The specific angular momentum (or, the angular momentum per unit mass) of the gyrating ellipsoid can be computed as, ${\bf{h}}={\bf{r}}\times{\bf{v}}$, where ${\bf{r}}$ and ${\bf{v}}$ are the position and velocity of the center-of-mass of the ellipsoid in 2D. This kinematic quantity, like angular velocity \cite{mancois2018two,dotsenko2013two,bae2021inertial}, also describes a planar rotation. It essentially measures the propensity of the ellipsoid to keep gyrating, about the origin of the lab frame, by always maintaining a finite mean distance between the origin and its center-of-mass. For a charged particle gyrating in a magnetic field \cite{muhsin2025active}, the specific angular momentum is proportional to the orbital magnetic moment. It also dictates the heat current during microscopic gyration \cite{bae2021inertial}. Solving Eq.[\ref{eom1},\ref{eom2}] numerically, we obtain the positions and velocities of the center-of-mass of the ellipsoid, and compute the average specific angular momentum $(\langle h \rangle)$ in the steady-state. In the following plots, the variation of $\langle h \rangle$ has been studied with respect to all the parameters relevant to our model.

\begin{figure}[htp]
    \centering
    \includegraphics[width=8cm]{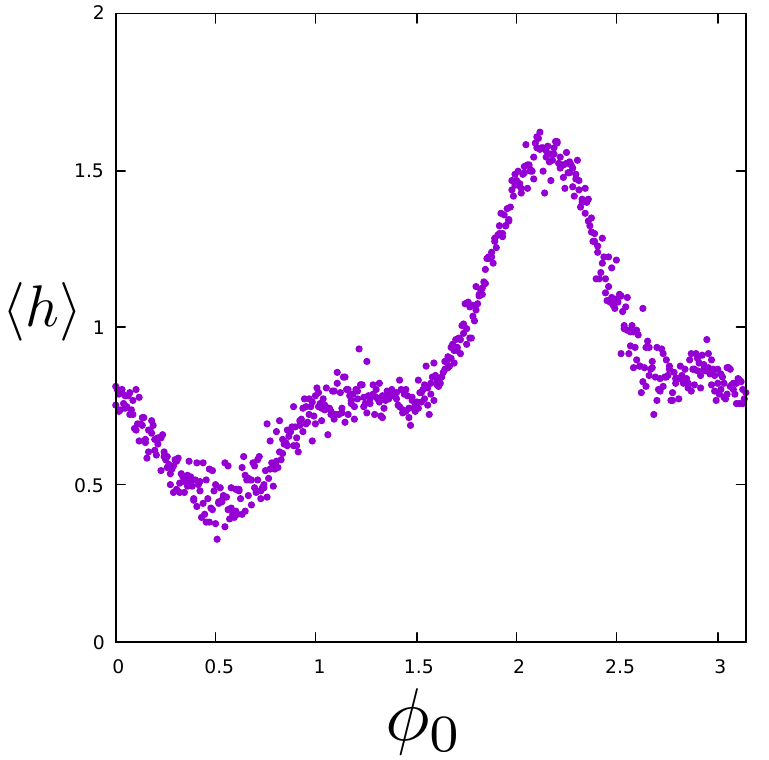}
    \caption{Plot of $\langle h \rangle$ versus $\phi_0$ (in radian, within the range $[0:\pi]$). The relevant values of mean orientation, namely $\phi_0=\frac{\pi}{4},\frac{\pi}{2},\frac{3\pi}{4}$, can be found as 0.78, 1.57 and 2.36 respectively on the $\phi_0-$axis. The values of the system parameters, such as $k_{\phi}, |\Delta\gamma|, \gamma_r$ etc. are given in Sec.III A.} 
\label{phi0}    
\end{figure}

Fig.[\ref{phi0}] shows the variation of $\langle h \rangle$ with the mean orientation of the ellipsoid $(\phi_0)$. From the figure it is evident that $\phi_0$ can indeed modulate the magnitude of $\langle h \rangle$. However, for a given set of $\alpha$, $T_x$ and $T_y$, this magnitude does not cross zero which implies that, by varying $\phi_0$, one cannot nullify the gyration or alter its handedness (i.e. the clockwise or anti-clockwise sense). From Eq.[\ref{eom1}], it is evident that the center-of-mass dynamics of the ellipsoid depends on the orientation angle $\phi$ of the ellipsoid, via the frictional drag coefficients. Consequently, $\langle h\rangle$ explicitly depends on $\langle \phi\rangle=\phi_0$. The net force coming from the frictional drag along $x$ (say) takes the following form: $F_x^{\text{drag}}\equiv -\gamma_{xx}v_x + \left(\frac{|\Delta\gamma|}{2}\sin{2\phi}\right)v_y$. The total drag along $x$ can thus be considered as a sum of two contributions, namely $F_x^{\text{principal}}=-\gamma_{xx}v_x$ and $F_x^{\text{off-diagonal}}=-\gamma_{xy}v_y=\left(\frac{|\Delta\gamma|}{2}\sin2\phi\right)v_y$ . Here, $\gamma_{xx}$ (in $F_x^{\text{principal}}$), though varies with $\phi$, but always remains positive. However, $\gamma_{xy}$ (that relates $\dot{v}_x$ with $v_y$), can flip its sign, depending on the orientation angle $\phi$ of the ellipsoid. Therefore, $F_x^{\text{off-diagonal}}$ can reduce $F_x^{\text{drag}}$, when $\gamma_{xy}< 0$. This eventually affects $\langle h\rangle$, when the average orientation angle of the ellipsoid $\phi_0$ varies.  A similar explanation is also applicable for the dynamics along $y$.


From Fig.[\ref{phi0}], it is evident that when $\phi_0=0$, the coupling between $v_x$ and $v_y$ vanishes on an average over all possible values of $\phi$ and $\langle h\rangle \simeq  0.8$ (note that $x$ and $y$ are always coupled for $\alpha\neq 0$, resulting in gyration).  As $\phi_0$ increases up to $\pi/4$, the value of $\langle \sin{2\phi}\rangle$ increases up to its maximum, which eventually maximizes the coupling between $v_x$ and $v_y$ and reduces $\langle h\rangle$ to its minimum (around $0.3$) . With further increase in $\phi_0$, $\langle\sin{2\phi}\rangle$ gradually decreases to zero again at around $\phi_0=\frac{\pi}{2}$. This reduces the velocity coupling and increases $\langle h \rangle$, which finally becomes stable at around $0.8$, when $\phi_0$ becomes close to $\pi/2$. With further increase in $\phi_0$, $\langle \sin{2\phi}\rangle$ increases up to its negative maximum at around $\phi_0=\frac{3\pi}{4}$. This increases the value of $\langle h \rangle$ from around $0.8$ till it reaches around $1.5$. Then, upon further increase of $\phi_0$ from $3\pi/4$ to $2\pi$, $\langle \sin2\phi\rangle$ as well as the average velocity coupling decreases and finally vanishes. Hence, $\langle h\rangle$ also decreases till it reaches around $0.8$ again. The entire variation then gets repeated periodically. Thus, by carefully tuning $\phi_0$  and keeping other system parameters ($\alpha, T_x, T_y$, etc.) fixed, one can modulate the value of $\langle h \rangle$.

\begin{figure}[htp]
    \centering
    \includegraphics[width=8cm]{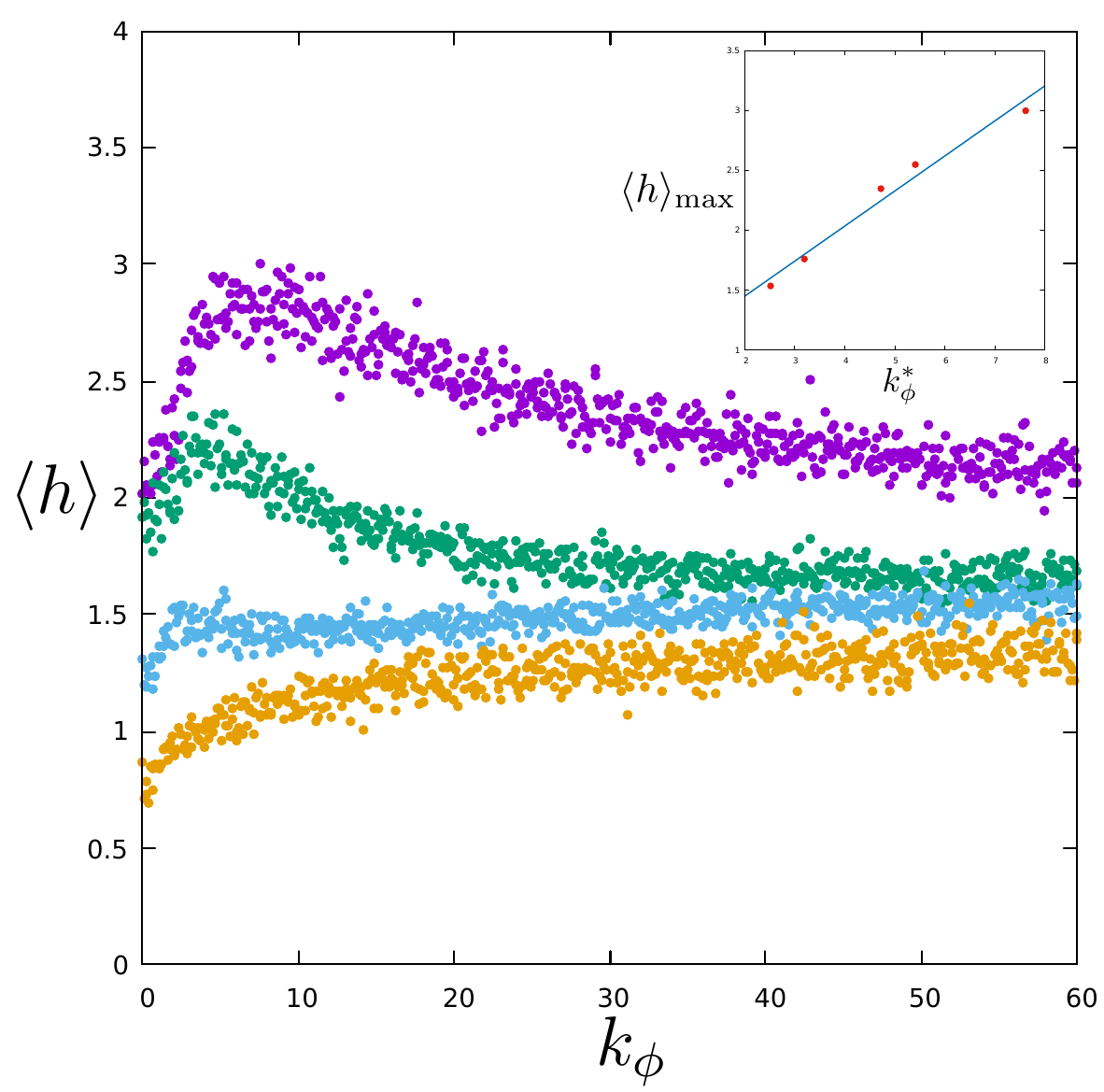}
    \caption{Plot of $\langle h \rangle$ versus $k_\phi$, for various masses : the purple, green, blue and orange points are for $m=0.01,0.1,1,10$ respectively. The inset shows the variation of $\langle h \rangle_\text{max}$ versus $k_\phi^*$ for five different masses, $m=0.01,0.05,0.1,0.5,1$ (where the peaks in $\langle h \rangle$ are prominent). The slope of the best fit line $(y=0.292x+0.866)$ gives an estimate of the dynamical susceptibility.}
\label{kphi}    
\end{figure}

Fig.[\ref{kphi}] shows the variation of $\langle h \rangle$ with the external torque strength, $k_\phi$, for different values of $m$. When the restoring torque on the ellipsoid is zero (i.e. when $k_{\phi}=0$ and the orientation of the ellipsoid is freely-diffusing), the value of $\langle h \rangle$ is inversely proportional to the mass of the ellipsoid which implies that inertia inhibits gyration. As $k_{\phi}$ increases from zero, $\langle h\rangle$ starts to increase too. However, for heavier particles, it monotonically saturates for larger values of $k_{\phi}$, without exhibiting any prominent maxima. On the other hand, for small values of $m$, $\langle h\rangle$ varies non-monotonically with increasing $k_{\phi}$. In this case, as $k_{\phi}$ increases from zero, the value of $\langle h \rangle$ becomes maximum at a certain $k_\phi=k_\phi^*$ and then, upon increasing $k_{\phi}$  beyond $k_{\phi}^*$, $\langle h\rangle$ gradually decreases. Finally, it saturates for larger values of $k_{\phi}$. The peak value of $\langle h \rangle$ ($\equiv \langle h\rangle_{\text{max}}$, say) increases with decreasing $m$. This implies that in the limit of $m\rightarrow0$ (i.e. the overdamped limit), as a microscopic gyrator, the Brownian ellipsoid becomes more susceptible to the restoring torque around $k_{\phi}^*$. To quantify the \enquote{susceptibility} of the gyrator towards an external torque, $\langle h \rangle_\text{max}$ versus $k_\phi^*$ can be plotted (see the inset of Fig.[\ref{kphi}]). From the plot, one may note that for lower values of $m$, $\langle h \rangle_\text{max}$ occurs at slightly higher values of $k_\phi^*$. Furthermore, a straight line can be fitted with $\langle h \rangle_\text{max}$ versus $k_\phi^*$ plot, the slope of which provides an estimate of how the gyrating ellipsoid \enquote{responds} to the torque that tries to restore its orientation towards $\phi_0$. Thus, the slope $\chi_\text{torque}=\frac{\Delta\langle h \rangle_\text{max}}{\Delta k_\phi^*}$, can be termed as the \emph{dynamical susceptibility} of the inertial ellipsoid towards a restoring torque in the midst of its gyration, which can serve as a characteristic feature of the system. As $k_\phi$ is increased to larger values beyond $k_{\phi}^*$, the fluctuations in $\phi$ are reduced, which eventually reduces the fluctuations in $h$. Hence, a higher torque assists the ellipsoid in stabilizing its gyratory response.

\begin{figure}[htp]
    \centering
    \includegraphics[width=8cm]{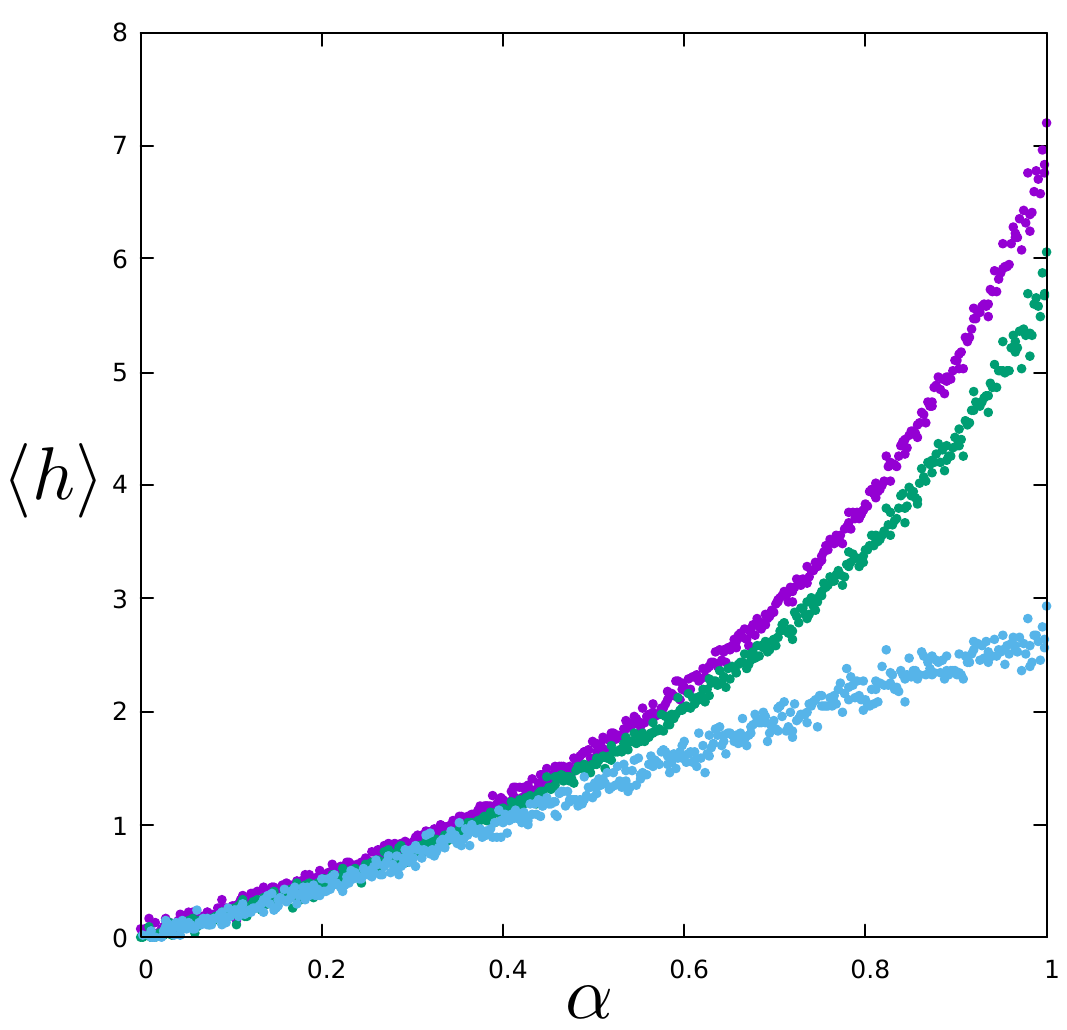}
    \caption{Plot of $\langle h \rangle$ versus $\alpha$ (with $\alpha>0$, and $\alpha\leq k=1$), for various masses : the purple, green and blue points are for $m=0.1,1,10$ respectively.}
\label{alpha}    
\end{figure}

Fig.[\ref{alpha}] shows the influence of the asymmetry present in the trapping potential on $\langle h \rangle$, for different values of $m$. For a spherically-symmetric harmonic potential $(\alpha=0)$, the ellipsoid fails to connect the hot and cold baths and the NESS is absent. Consequently, gyration is also absent which leads to a zero angular momentum. For lower values of $\alpha$, the variation is linear for all masses $(\langle h \rangle \sim \alpha)$ and there is no significantly distinguishable effect of inertia. However, as $\alpha$ increases further, the inertial effect becomes prominent --- the qualitative feature of $\langle h\rangle$ vs. $\alpha$ plots becomes different for different values of $m$. Due to the inertial inhibition, the ellipsoid with a higher mass shows a gradual, almost linear increase in $\langle h \rangle$ with $\alpha$. However, lower values of masses show a rapid, non-linear increment of $\langle h \rangle$ with increasing $\alpha$. Our results agree with the physical reality described in \cite{bae2021inertial} that the rapid variation of angular momentum with potential asymmetry subsides with an increasing mass, due to an increase in the moment of inertia of the particle which makes the gyration difficult to perform.


However, apart from the similarity, here we notice a major difference between $\langle h\rangle$ vs. $\alpha$ plots for spherically-symmetric and highly asymmetric cases with large $|\Delta\gamma|$. The  variation of $\langle h \rangle$ with $\alpha$ as observed for a spherical gyrator with finite inertia is non-monotonic, and therefore, $\langle h\rangle$ becomes maximum at a certain $\alpha=\alpha^*$ \cite{bae2021inertial}. On the other hand, here $\langle h\rangle$ increases monotonically with $\alpha$ and no particular value of $\alpha$ is favored by the inertial ellipsoid as far as its gyratory response is tested with varying $\alpha$. The decrease in $\langle h \rangle$ at higher $\alpha$, as observed in a gyrating inertial, spherical particle \cite{bae2021inertial}, is prevented by a non-zero, large $|\Delta\gamma|$, that is, by the spherical asymmetry present in the shape of the ellipsoid. Whether the decrease in angular momentum can be recovered in the small $|\Delta\gamma|$ limit will be verified later with numerical as well as analytical results.


\begin{figure}[htp]
    \centering
    \includegraphics[width=8cm]{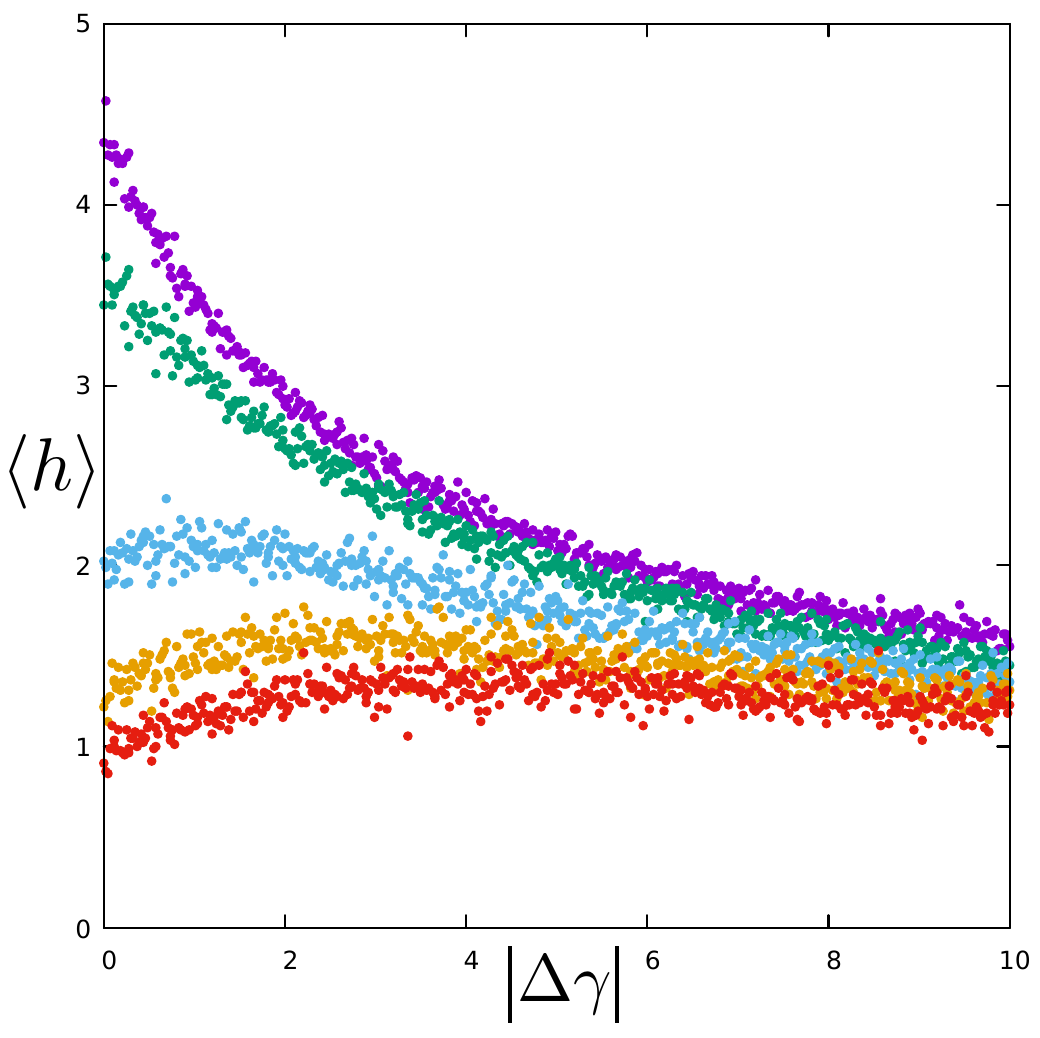}
    \caption{Plot of $\langle h \rangle$ versus $|\Delta\gamma|$, for various masses : the purple, green, blue, orange and red points are for $m=0.1,1,5,10,15$ respectively.}
\label{deltagamma}    
\end{figure}

The effects of the spherical asymmetry of the shape of the ellipsoid on its gyration are depicted in Fig.[\ref{deltagamma}]. The spherical limit is indicated by $|\Delta\gamma|=0$ and the values of $\langle h \rangle$ agree with the analytical result derived in the subsequent section. Note that $\alpha (\neq0)$, as expected, facilitates the gyration to survive even in the spherical limit. As discussed earlier and in \cite{bae2021inertial}, in the spherical limit, the value of $\langle h \rangle$ decreases with an increasing mass. For lower masses, the angular momentum monotonically decreases with increasing spherical asymmetry (i.e. increasing $|\Delta\gamma|$). This  indicates that lighter ellipsoids tend to inhibit their gyration owing to their shape. Due to a non-zero $|\Delta\gamma|$, the planar velocity components ($v_x$ and $v_y$) of the ellipsoid get coupled, which effectively makes it difficult for the ellipsoid to gyrate. As the mass is increased, the variation alters qualitatively --- $\langle h \rangle$ now varies non-monotonically with $|\Delta\gamma|$, where it slightly increases from the isotropic value and reaches a peak with increasing $|\Delta\gamma|$. While increasing $|\Delta\gamma|$ further, $\langle h\rangle$ decreases slowly. However, it is evident from Fig.[\ref{deltagamma}] that, with an increasing mass of the ellipsoid, the dependence of $\langle h\rangle$ on $|\Delta\gamma|$ becomes less pronounced and gyration almost becomes independent of the shape asymmetry.



\begin{figure}[htp]
    \centering
    \includegraphics[width=8cm]{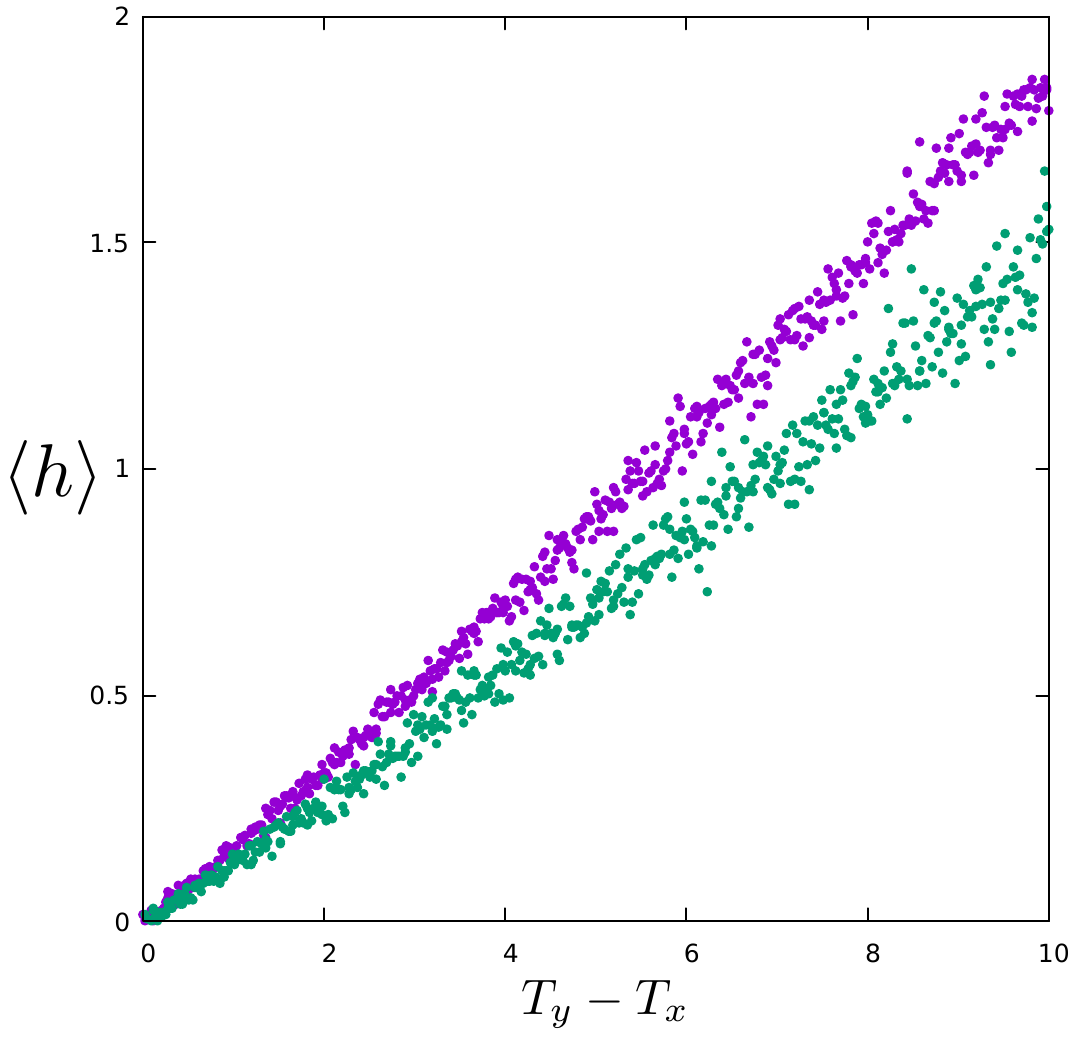}
    \caption{Plot of $\langle h \rangle$ versus $T_y-T_x$ (with $T_y>T_x$), for various masses : the purple and green points are for $m=0.1,10$ respectively.}
\label{tempdiff}    
\end{figure}

Fig.[\ref{tempdiff}] shows that $\langle h\rangle$ increases linearly with $T_y-T_x$ and its slope is higher for a lower mass of the ellipsoid.  If  $T_y-T_x$ is zero, $\langle h\rangle$ becomes zero, i.e. the ellipsoid cannot gyrate in the absence of a NESS.  The asymmetric trap and anisotropic shape connect the two orthogonal heat baths and cause the ellipsoid to gyrate in between these baths. The current, thus generated, is solely due to the non-equilibrium, steady-state heat flux that flows between the hot and cold baths (as decided by the inequality in temperatures). Whether the gyration occurs in a clockwise or an anti-clockwise sense will be decided by the sign of the quantity $\alpha(T_y-T_x)$, as will be discussed while deriving the analytical expression for $\langle h \rangle$ in the small-coupling limit.


\begin{figure}[htp]
    \centering
    \includegraphics[width=8cm]{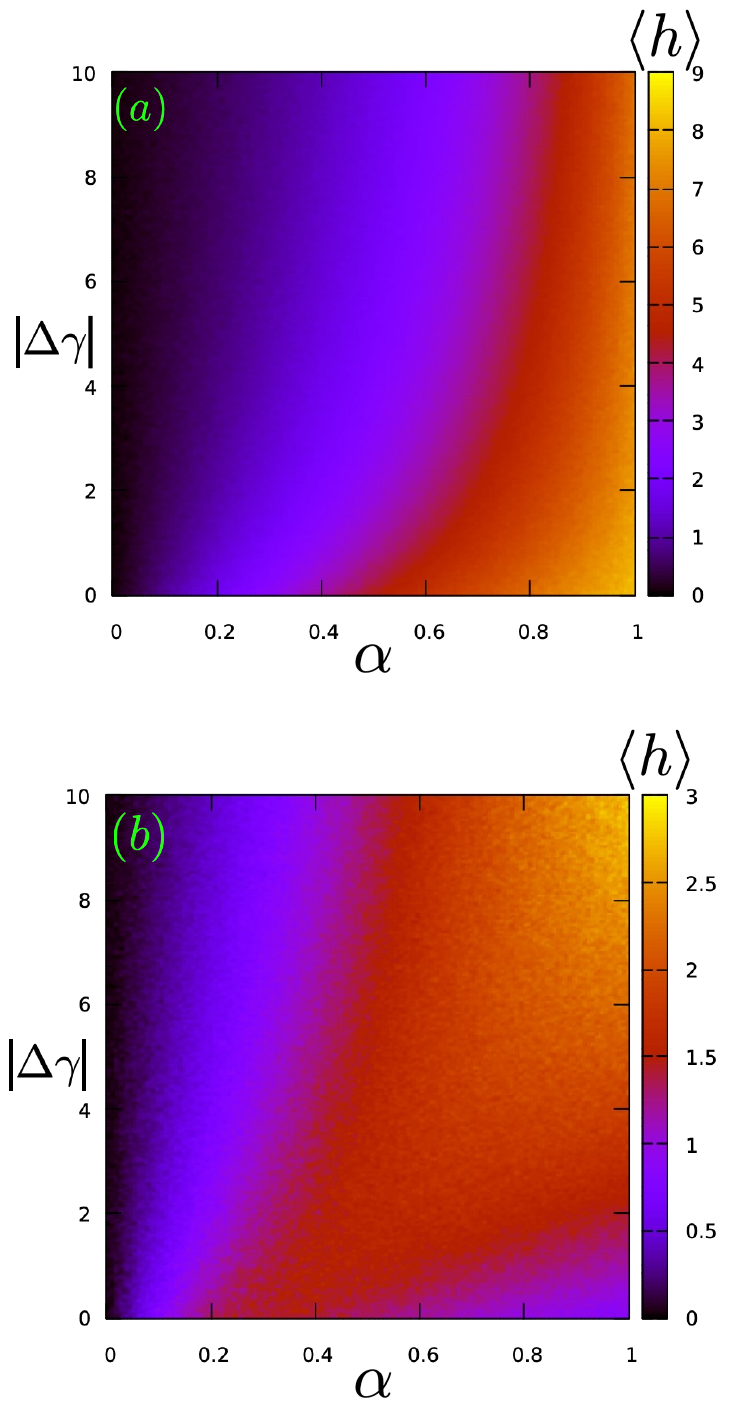}
    \caption{Parametric plots of $\langle h\rangle$ with $\alpha$ along $x-$axis, $|\Delta\gamma|$ along $y-$axis, and $\langle h \rangle$ as color, for two different masses --- (a) $m=0.1$, (b) $m=10$.}
\label{phasediagram}    
\end{figure}

Having discussed the variation of specific angular momentum with the relevant system parameters, it will be useful to study the simultaneous effects of both the asymmetries --- of shape and trap --- on the gyrating ellipsoid, and the role of mass therein. The parametric plots of $\langle h\rangle$ given in Fig.[\ref{phasediagram}] will essentially capture the interplay of inertia and anisotropy. The $\alpha=0$ boundary denotes the case where no gyration occurs. For lower inertia, along the $|\Delta\gamma|=0$ boundary (i.e. spherical case), $\langle h\rangle$ increases with $\alpha$. As one goes to higher values of shape anisotropy, lower values of $\langle h\rangle$ dominate the parameter-space. For higher mass, however, inertia of the ellipsoid suppresses the gyration, besides shape. Along  $|\Delta\gamma|=0$ boundary, $\langle h\rangle$ shows a non-monotonic variation with $\alpha$. For higher potential asymmetry (close to $k=1$), gyration becomes difficult to perform due to a large moment of inertia and $\langle h\rangle$ decreases as a result \cite{bae2021inertial}. However, this non-monotonicity is lost for increasing values of $|\Delta\gamma|$. 

In the next section, we will characterize the output of the gyrator in terms of the ratio of mean and fluctuation of $h$. We know that, in case of microscopic gyration, the heat flux from one bath to the other drives the system away from equilibrium, facilitating the system to gyrate, both in underdamped \cite{bae2021inertial} and overdamped cases \cite{filliger2007brownian}. The output can also be characterized in terms of this heat flux. However, here we will consider $h$ and its moments for the characterization. Average heat flux and its dependence on shape will be addressed in Appendix-II, where we have discussed the energetics of our ellipsoidal gyrator.  

\subsection{Output characteristics of the gyrator}

\begin{figure*}[htp]
    \centering
        \includegraphics[width=0.9\textwidth]{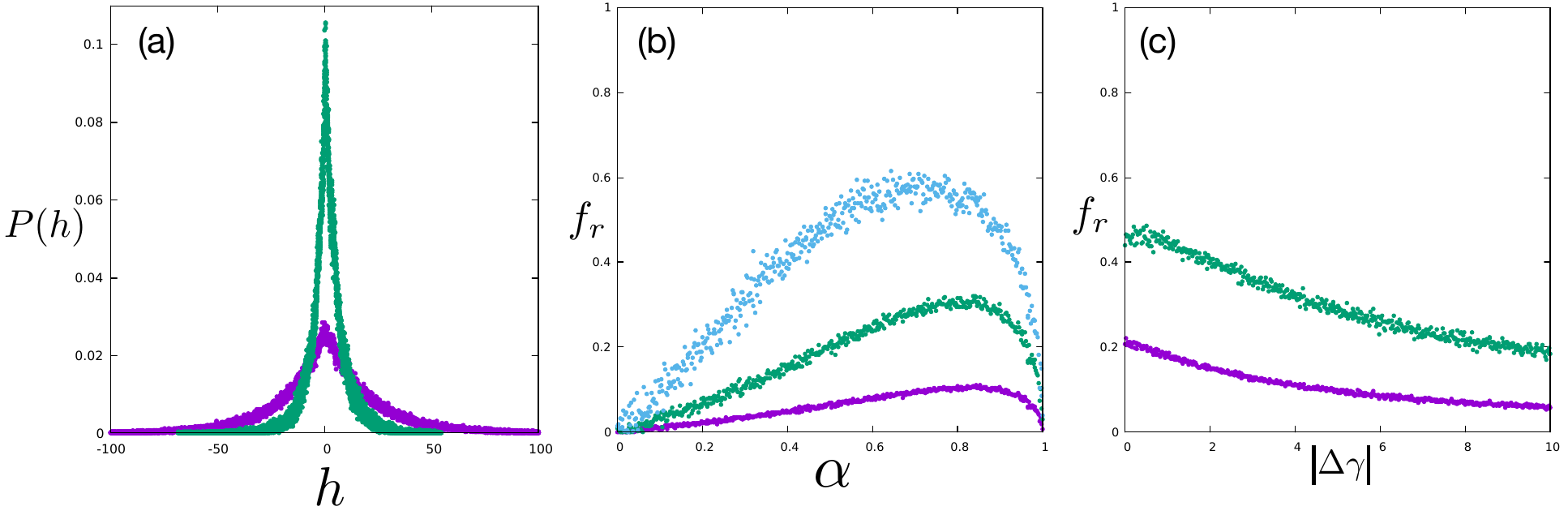}
    \caption{(a) Distribution of specific angular momentum : $P(h)$ versus $h$, for $m=0.1$ (purple) and $m=1$ (green). (b) Plot of $f_r$ versus $\alpha$, for various masses : the purple, green and blue points are for $m=0.1,1,10$ respectively. (c) Plot of $f_r$ versus $|\Delta\gamma|$, for various masses : the purple and green points are for $m=0.1,1$ respectively.}
\label{distri}    
\end{figure*}



\begin{figure*}[htp]
    \centering
    \includegraphics[width=0.9\textwidth,height=4cm]{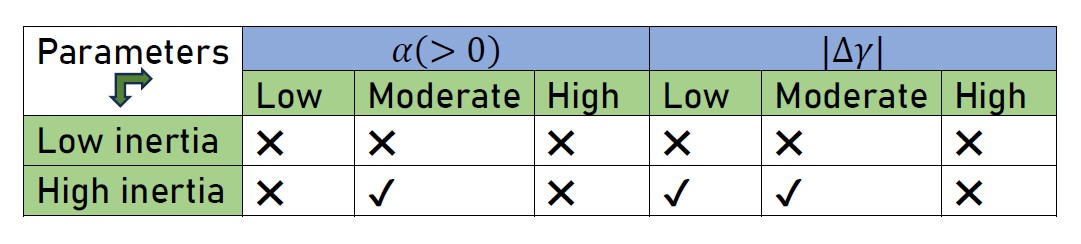}
    \caption{Summary of various performance regimes and the set of corresponding control parameters (mass, potential asymmetry, shape anisotropy) --- \enquote{tick} denotes a beneficial combination (high $f_r$), while \enquote{cross} denotes a detrimental combination (low $f_r$).}
\label{table}    
\end{figure*}

Here we will judge how \enquote{fine} the inertial ellipsoid gyrates, by exploring the steady-state distribution of specific angular momentum, for various values of $m$ --- as given in Fig.[\ref{distri} (a)]. A cusp-like behavior is observed at $h=0$. This implies that non-zero values of $h$ are supported by atypical trajectories, residing along the tails of the distribution. One may also note the effect of inertia on the width of the distribution --- a higher mass suppresses the fluctuations in angular momentum. We will find here that this is indeed beneficial for the microscopic gyration.

Fluctuations can have an inhibitory effect on Brownian gyration in the overdamped regime \cite{viot2023destructive}. We need to identify the parameter regimes where stochastic fluctuations become detrimental --- the primary aim of improving gyration lies in reducing the fluctuations in angular momentum as compared to its mean value. It can be done by suitably tuning the so-called \emph{signal-to-noise ratio}, $f_r\equiv\frac{\langle h \rangle}{\sigma_h}$, i.e. the ratio of mean angular momentum to its standard deviation $(\sigma_h = \sqrt{\langle h^2 \rangle - \langle h \rangle^2})$, where the latter quantifies the fluctuations of angular momentum around its mean.

Figs.[\ref{distri} (b),(c)] show the effects of $\alpha$ and $|\Delta\gamma|$  on $f_r$ and how inertia can facilitate a better gyration. Both the plots show that $f_r$ is always less than unity, indicating the significance of fluctuations. In Fig.[\ref{distri} (b)], $f_r$ goes to zero due to the absence of gyration in $\alpha\rightarrow 0$ limit. In this limit, the spherically-asymmetric harmonic trap tends to become spherically-symmetric, and therefore, $\langle h\rangle\rightarrow 0$. As $\alpha$ increases from zero,  Fig.[\ref{alpha}] suggests that $\langle h\rangle$ also increases, facilitating the system to gyrate, and therefore, $f_r$ increases. However, $f_r$ cannot grow with $\alpha$ forever.  As $\alpha$ increases from zero to $k$ (with $k$ being fixed), it becomes increasingly difficult to keep the particle bounded within the trap. Therefore, the position and velocity fluctuations and consequently $\sigma_h$ of the particle become large. This makes $f_r\rightarrow 0$, as $\alpha\rightarrow k$. 


Between these two extremes ($\alpha\rightarrow 0$ and $\alpha\rightarrow k$), we find an important non-monotonic variation of $f_r$, where it is maximized at some $f_r\equiv f_r^*$ , at a particular $\alpha=\alpha^*$. The value of this optimum potential asymmetry $(\alpha^*)$ slightly shifts towards $k$ as $m$ decreases (see Fig.[\ref{distri} (b)]). More importantly, as the fluctuations are lower for higher values of $m$ (i.e. for larger inertia), $f_r^*$ increases with increasing $m$ of the gyrating ellipsoid. This reveals the fact that inertia can facilitate microscopic gyration of a Brownian ellipsoid by enhancing the signal-to-noise ratio, with  $0<\alpha< k$ . To obtain the maximal change in $f_r$ by varying $m$, it would be best to reside in the periphery of $\alpha=\alpha^*$.       

In Fig.[\ref{distri} (c)], on the other hand, the value of $f_r$ monotonically decreases with $|\Delta\gamma|$ because of decreasing $\langle h\rangle$ and increasing $\sigma_h$ (see Fig.[\ref{deltagamma}]). Due to a more prominent inertial contribution, the values of $f_r$ remain sufficiently high for higher mass.

A summary of all the operating regimes, both beneficial and detrimental, of the gyrator has been given in Fig.[\ref{table}]. This should be useful to obtain an optimized microscopic gyration from the inertial ellipsoid, by suitably tuning the major control parameters.



\section{Analytical results for the slightly anisotropic system}

\subsection{Approximations and linearized equations of motion}

According to Eq.[\ref{eom1}],  the coupling between $v_x$ and $v_y$  survives statistically if $\langle\sin\phi\cos\phi\rangle=\int \sin\phi\cos\phi P(\phi) d\phi \neq 0$, where $P(\phi)$ is the steady-state distribution of $\phi$ and the integral is taken over all possible values of $\phi$. Clearly, the distribution must have a broken inversion symmetry w.r.t. the angle $\phi$, that is, $P(\phi)\neq P(-\phi)$. This can be achieved by employing an external torque, as introduced in Eq.[\ref{eom2}] --- the net torque fluctuates in time, such that $\langle\sin\phi\cos\phi\rangle\neq0$. The non-zero value of $\phi_0$ helps in breaking the inversion symmetry of the corresponding distribution of angular orientation about $\phi=0$. If the steady-state orientational distribution is asymmetric, it will affect the planar translation of the center-of-mass of the ellipsoid by coupling its orthogonal velocity components. Keeping this in mind, two physically-motivated approximations can be used to achieve the limit of small shape anisotropy, which can be exhibited by an ellipsoid that slightly deviates from a perfect spherical symmetry:

(i) One can consider a restoring torque, in the dynamics of $\phi$, with a large value of the torque-strength $k_\phi$ and a small mean orientation $\phi_0=\delta\phi$, i.e. the small angle to which the orientation is relaxed at large times. Large value of $k_\phi$ suppresses the angular fluctuations, leading to a faster relaxation of the orientational DoF to its mean value. Hence, the translational dynamics of the ellipsoid can be averaged over all possible values of $\phi$. Such a suppression has another obvious consequence --- the steady-state variance of $\phi$ (that is, the \enquote{width} of the distribution $P(\phi)$) becomes small. Thus, the first-order moment of $\phi$ (i.e. $\delta\phi$) has the dominant contribution, and the second and other higher-order moments are quite small in comparison to the first moment, while the averaging of $\sin\phi\cos\phi$ is done across all possible $\phi$. 

(ii) One can further simplify the dynamical equations by taking the limit $|\Delta\gamma|\longrightarrow 0$, but not exactly equal to zero \cite{mandal2024diffusion}. In this limit, $\gamma_{\perp}$ and $\gamma_{\parallel}$ are so close to each other that both can be replaced by their average, $\gamma=\frac{\gamma_{\perp}+\gamma_{\parallel}}{2}$. The limit is valid for particles having a shape which slightly deviates from a perfectly isotropic shape. This can be practically achieved by tuning the {\it{aspect ratio}} ($\mathcal{A}$) \cite{dhont1996introduction,berg1993random} of the ellipsoidal particle, which is simply the ratio of the lengths of its long and short principal axes. Aspect ratio can also be thought of as a measure of the shape anisotropy of the particle, where, $\mathcal{A}>1$ denotes a prolate ellipsoid \cite{berg1993random}. Thus, the value of $\mathcal{A}$ can be suitably altered for fabricating an ellipsoid of a definite shape.

Using these approximations and taking an average on both sides of Eq.[\ref{eom1}] w.r.t. $P(\phi)$ (that is, integrated over all possible values of $\phi$),  one can now obtain the following set of linear differential equations governing the simplified, $\phi$-averaged, effective translational dynamics of the ellipsoid with non-negligible inertia:

\begin{eqnarray}
\nonumber
m(\partial_t v_x)=-\gamma(v_x - |\epsilon|v_y)-\frac{\partial U(x,y)}{\partial x}+\sqrt{2\gamma T_x}\zeta_x \\
m(\partial_t v_y)=-\gamma(v_y - |\epsilon|v_x)-\frac{\partial U(x,y)}{\partial y}+\sqrt{2\gamma T_y}\zeta_y 
\label{eom3}
\end{eqnarray}

where, $\epsilon\equiv\left(\frac{\Delta\gamma}{\gamma}\right)\delta\phi$ is the dissipative coupling parameter, and, $\langle \zeta_i (t)\rangle=0$, $\langle\zeta_i(t)\zeta_j(t')\rangle=\delta_{ij}\delta(t-t')$. The coupling parameter is negative and its magnitude is much less than unity. It is evident that, even in the limit of small anisotropy, the coupling between $v_x$ and $v_y$ survives in Eq.[\ref{eom3}] through the geometric anisotropy $(|\Delta\gamma|)$ of the ellipsoid and its non-zero, average orientation $(\delta\phi)$, both being intrinsic features of the particle itself. Needless to say that the above set of coupled differential equations will also lead to a steady-state gyration, as evident from its generic form. We emphasize here that these approximations are not mandatory to obtain gyration --- the simplification turns the problem into an analytically-tractable one.

In the following few sub-sections, we will discuss the method to obtain an expression for the average, steady-state specific angular momentum, using Eq.[\ref{eom3}].

\subsection{Steady-state covariance matrix }


In our case, the covariance matrix $\Sigma_{\mu\nu}\equiv\langle \mu \nu \rangle - \langle \mu \rangle \langle \nu \rangle$ with $(\mu,\nu)\in\{x,y,v_x,v_y\}$, obeys the following equation in steady-state \cite{gardiner1985handbook,van1992stochastic,mancois2018two}:

\begin{equation}
    A\Sigma+ \Sigma A^T + B=0
\label{lyapunov}
\end{equation}

where, $A$ and $B$ are the matrices containing system parameters. These $4\times4$ matrices can be explicitly defined as: 

\begin{widetext}

\begin{eqnarray}
    \Sigma = \begin{bmatrix} \langle x^2 \rangle & \langle xy \rangle & \langle xv_x \rangle & \langle xv_y \rangle \\ \langle yx \rangle & \langle y^2 \rangle & \langle yv_x \rangle & \langle yv_y \rangle \\ \langle v_x x\rangle & \langle v_x y \rangle & \langle v_x^2 \rangle & \langle v_x v_y \rangle \\ \langle v_y x \rangle & \langle v_y y \rangle & \langle v_y v_x \rangle & \langle v_y^2 \rangle \end{bmatrix}
\end{eqnarray}

\begin{eqnarray}
A = \begin{bmatrix} 0 & 0 & 1 & 0 \\ 0 & 0 & 0 & 1 \\-\frac{k}{m} & -\frac{\alpha}{m} & -\frac{\gamma}{m} & \frac{|\epsilon|\gamma}{m} \\ -\frac{\alpha}{m} & -\frac{k}{m} &\frac{|\epsilon|\gamma}{m} & -\frac{\gamma}{m} \end{bmatrix}
\end{eqnarray}

\begin{eqnarray}
B = \begin{bmatrix} 0 & 0 & 0 & 0 \\ 0 & 0 & 0 & 0 \\ 0 & 0 & \frac{2\gamma T_x}{m^2} & 0 \\ 0 & 0 & 0 & \frac{2\gamma T_y}{m^2} \end{bmatrix}
\end{eqnarray}
    
\end{widetext}

Plugging the system matrices in Eq.[\ref{lyapunov}], all the steady-state correlations corresponding to translational dynamics of the center-of-mass motion of the ellipsoid (in the small-coupling limit) can be explicitly calculated in terms of the system parameters:

\begin{widetext}

\begin{eqnarray}
\nonumber
\langle x^2 \rangle &=& \frac{(T_x+T_y)(k-\alpha|\epsilon|)}{2(k^2-\alpha^2)(1-\epsilon^2)}-\frac{\gamma^2(T_y-T_x)}{2[\gamma^2(k+\alpha|\epsilon|)+m\alpha^2]}\\
\nonumber
\langle xy \rangle (= \langle yx \rangle) &=& \frac{(T_x+T_y)(k|\epsilon|-\alpha)}{2(k^2-\alpha^2)(1-\epsilon^2)}\\
\nonumber
\langle xv_x \rangle (= \langle v_x x \rangle) &=& 0\\
\nonumber
\langle xv_y \rangle (= \langle v_y x \rangle) &=& \frac{\gamma\alpha(T_y-T_x)}{2[\gamma^2(k+\alpha|\epsilon|)+m\alpha^2]}\\
\nonumber
\langle y^2 \rangle &=& \frac{(T_x+T_y)(k-\alpha|\epsilon|)}{2(k^2-\alpha^2)(1-\epsilon^2)}-\frac{\gamma^2(T_x-T_y)}{2[\gamma^2(k+\alpha|\epsilon|)+m\alpha^2]}\\ 
\nonumber
\langle y v_x \rangle (= \langle v_x y \rangle) &=& -\langle xv_y \rangle\\
\nonumber
\langle yv_y \rangle (= \langle v_y y \rangle) &=& 0\\
\nonumber
\langle v_x^2 \rangle &=& \frac{T_x}{m}+\frac{\epsilon^2(T_x+T_y)}{2m(1-\epsilon^2)}+\frac{\alpha^2(T_y - T_x)}{2[\gamma^2(k+\alpha|\epsilon|)+m\alpha^2]}\\
\nonumber
\langle v_x v_y \rangle (= \langle v_y v_x \rangle) &=& \frac{|\epsilon|(T_x+T_y)}{2m(1-\epsilon^2)}\\
\langle v_y^2 \rangle &=& \frac{T_y}{m}+\frac{\epsilon^2(T_x+T_y)}{2m(1-\epsilon^2)}+\frac{\alpha^2(T_x - T_y)}{2[\gamma^2(k+\alpha|\epsilon|)+m\alpha^2]}
\label{covariance}
\end{eqnarray}

\end{widetext}

The relations given in Eq.[\ref{covariance}] require some elucidation. As $|\epsilon|<1$, there is no possibility of the moments being divergent at $|\epsilon|\rightarrow1$.  For $|\epsilon|\geq 1$ these relations are not valid simply because the approximations do not hold. The correlation $\langle xy \rangle$ vanishes only for $|\epsilon|=0=\alpha$, or when $|\epsilon|=\frac{\alpha}{k}$. Also, it shows negative values only $(k|\epsilon|<\alpha)$ and diverges when $\alpha\rightarrow k$. Similar variations have been observed in Fig.[\ref{xyavg}]. At equilibrium and without any coupling whatsoever, the mean-squared velocities obey equi-partition theorem. The mean-squared positions also show consistent values in such a scenario. The coupling of velocities survives only in the anisotropic case $(|\epsilon|\neq0)$. For the spherical limit, all the moments in Eq.[\ref{covariance}] match the results reported in \cite{bae2021inertial,mancois2018two}. Using the relevant moments, we can now proceed towards calculating the mean-squared displacement of the center-of-mass of the ellipsoid at steady-state, along with the specific angular momentum in the small-coupling limit.


For a small anisotropy in shape, the steady-state mean-squared displacements of the center-of-mass of the ellipsoid along $x$ and $y$, as shown in Eq.[\ref{covariance}], can be added to obtain the radial variance as,  

\begin{eqnarray}
\langle r^2 \rangle = \frac{(T_x+T_y)(k-\alpha|\epsilon|)}{(k^2 - \alpha^2)(1-\epsilon^2)}
\label{radialeq}
\end{eqnarray}

For a spherical particle, $|\epsilon|=0$, and we obtain $\langle r^2 \rangle_{\text{spherical}} = \frac{k(T_x+T_y)}{k^2 - \alpha^2}$.  In both the cases, if $\Lambda\equiv\frac{\alpha}{k}\rightarrow 1$, the radial variance diverges as $ \sim\frac{1}{1-\Lambda^2}$.  This occurs due to the fact that, in this limit, the trapping potential takes the form $U(x,y)=\frac{k}{2}(x+y)^2$, and  the minima of this potential lies along the straight line $x+y=0$, thus leading to an unbounded motion. Hence, $\langle r^2 \rangle = \langle x^2 \rangle + \langle y^2 \rangle$ diverges, leading towards the detriment of gyration. 




\subsection{Specific angular momentum --- Reconciliation of numerical and analytical results} 

The average specific angular momentum in the steady-state can be calculated analytically in the small-coupling limit, using the cross velocity-position correlations given in Eq.[\ref{covariance}]:

\begin{eqnarray}
\langle h \rangle \equiv \langle xv_y \rangle - \langle yv_x \rangle = \frac{\gamma\alpha(T_y-T_x)}{\gamma^2(k+\alpha|\epsilon|)+m\alpha^2}
\label{h}
\end{eqnarray}

The above expression now confirms that the mean specific angular momentum is indeed inversely proportional to $|\Delta\gamma|$ as well as $m$. The spherical limit $(|\epsilon|=0)$ matches exactly with the numerical values shown in Fig.[\ref{deltagamma}] at $|\Delta\gamma|=0$ and also with \cite{bae2021inertial}. For a spherical particle in the overdamped limit, $\langle h\rangle$ can also be obtained from Eq.[\ref{h}] by setting $\epsilon\rightarrow0$ and $m\rightarrow0$ \cite{bae2021inertial}. It is evident from the above expression that gyration disappears when either potential asymmetry is absent or for a null temperature difference. The direction of gyration (clockwise or anti-clockwise) is solely determined by the collective sign of the product term $\alpha(T_y-T_x)$ \cite{muhsin2025active,bae2021inertial,dotsenko2013two,mancois2018two,filliger2007brownian}. For small $\alpha$, irrespective of whether $|\Delta\gamma|$ is large or small,  the angular momentum is proportional to $\alpha$. Hence, it is true for large $\Delta\gamma$ (see Fig.[\ref{alpha}]) and also for small $|\Delta\gamma|$, i.e. for small $|\epsilon|$ (as evident from Eq.[\ref{h}], with $\alpha\rightarrow 0$). As $\alpha$ increases, the inertial term $(\sim m\alpha^2)$ in the denominator starts to become significant and acquires distinct values for different masses, leading to a widely-separated variation of $\langle h\rangle$ with different $m$. This, as well as all other aspects of  Eq.[\ref{h}], will be verified by simulation. The fact that the effect of inertia becomes prominent for large $\alpha$ is true for both large $|\Delta\gamma|$ (see Fig.[\ref{alpha}]) and small $|\Delta\gamma|$ (as per Eq.[\ref{h}]). However, the large $|\Delta\gamma|$ behaviors of $\langle h\rangle$ cannot be captured by the formula in Eq.[\ref{h}], as it is derived within the premises of small $|\Delta\gamma|$, small $\phi_0$ and large $k_{\phi}$.

\begin{figure}[htp]
    \centering
    \includegraphics[width=7cm]{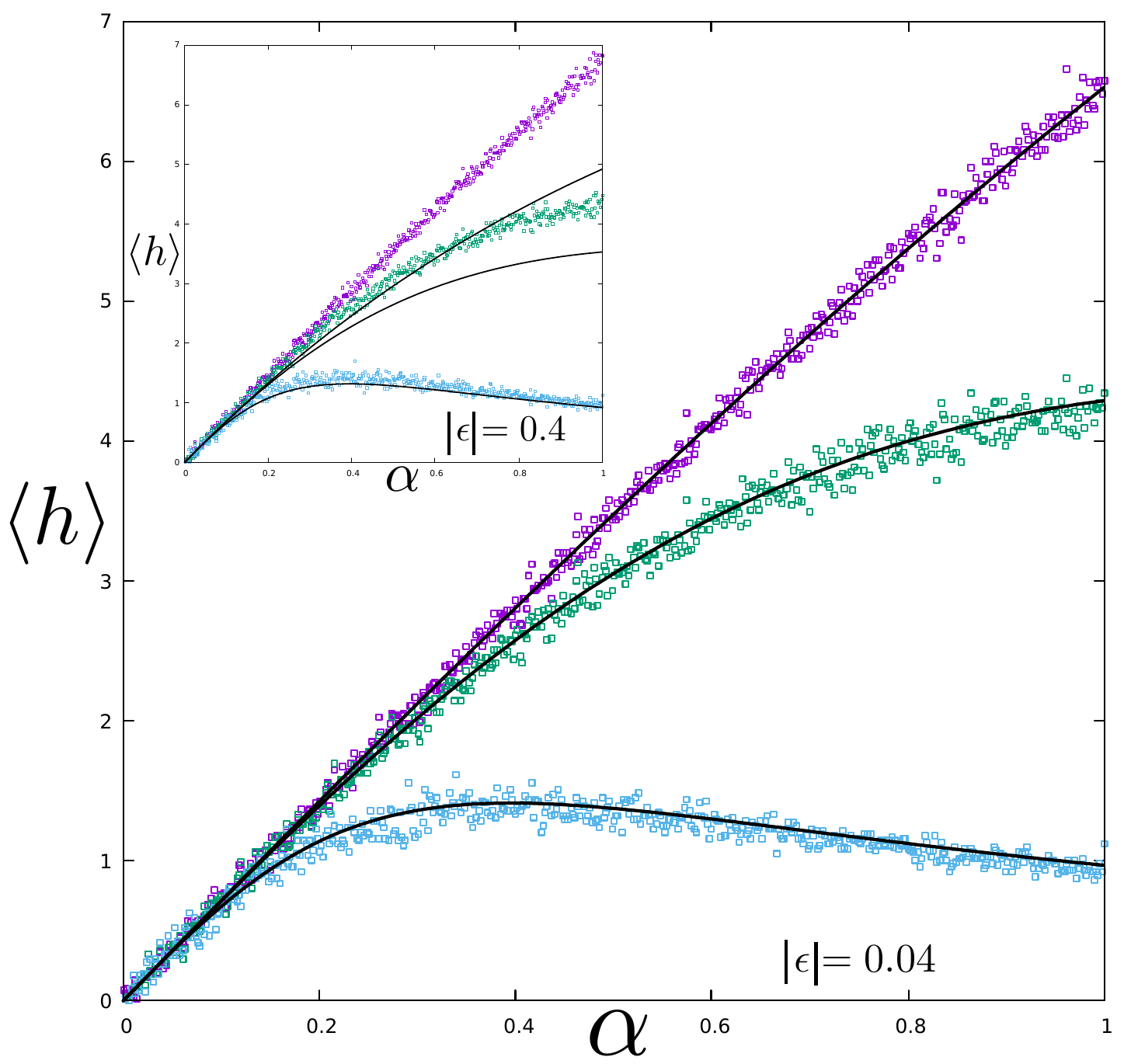}
    \caption{Plot of the numerical data of $\langle h \rangle$ versus $\alpha$, for $\gamma_\parallel=1$ and $\gamma_\perp=1.5$ : the main plot is for $\phi_0=0.1$ and $k_\phi=50$ (i.e. $|\epsilon|=0.04$) and the inset is for $\phi_0=1$ and $k_\phi=1$ (i.e. $|\epsilon|=0.4$). In both the plots, purple, green and blue points denote the data for $m=0.1,1,10$ respectively. The solid black lines are the respective analytical expressions given by Eq.[\ref{h}]. The inset clearly shows a poor agreement, due to large $\phi_0$ and small $k_\phi$.}
\label{reconcile}    
\end{figure}

We can now test the validity of our approximations by verifying whether the numerical results of the fully anisotropic system match with the analytical expression derived in the small-coupling limit, when the former is subjected to the stipulated range of parameters, that is --- small $|\Delta\gamma|$, small $\phi_0$ and large $k_\phi$ (see Fig.[\ref{reconcile}]). It is evident from Fig.[\ref{reconcile}] that, for sufficiently small $|\epsilon|$ (e.g. $|\epsilon|=0.04$), the analytical and numerical results match quite well. However, as expected, when $|\epsilon|$ becomes large (e.g. $|\epsilon|=0.4$), the disagreement becomes prominent. 

An important feature of the variation of $\langle h\rangle$ with respect to $\alpha$, with small anisotropy (i.e. small $|\epsilon|$) and large inertia (i.e. large $m$), is  the non-monotonicity --- it increases for small $\alpha$ and reaches up to a maximum and then decreases when $\alpha$ increases further. It becomes evident while comparing Fig.[\ref{alpha}] and Fig.[\ref{reconcile}] that this feature clearly differentiates the $\langle h\rangle$-vs.-$\alpha$ variation at large and small $|\epsilon|$, where the former variation is monotonic.  This upholds the role of shape anisotropy behind the microscopic gyration. From Eq.[\ref{h}], it can be shown that the peak of angular momentum occurs at $\alpha^* = \gamma\sqrt{\frac{k}{m}}$. For  $\gamma_\parallel=1$, $\gamma_\perp=1.5$, $k=1$ and $m=10$, we obtain $\alpha^*=0.395$, which exactly matches the value depicted in Fig.[\ref{reconcile}]. Also, one may note that the fractional change in $\langle h\rangle$ between the spherical and slightly non-spherical gyrating particles is given by: $\frac{\langle h\rangle_{|\epsilon=0|}-\langle h\rangle}{\langle h\rangle_{|\epsilon=0|}}=\frac{\gamma^2 \alpha |\epsilon|}{\gamma^2 (k+\alpha|\epsilon|)+m\alpha^2}$, which vanishes as $m\rightarrow \infty$. In this limit, $\langle h\rangle$ for spherical and slightly non-spherical cases are equal and independent of $|\epsilon|$. This is also reflected in Fig.[\ref{reconcile}], where the $\langle h\rangle$-vs.-$\alpha$ plots for $|\epsilon|=0.4$ and $|\epsilon|=0.04$ are quite close to each other for $m=10$.

\section{Manifestation of microscopic gyration - Anisotropic \enquote{magneto-gyrator}}

\begin{figure}[htp]
    \centering
    \includegraphics[width=8cm]{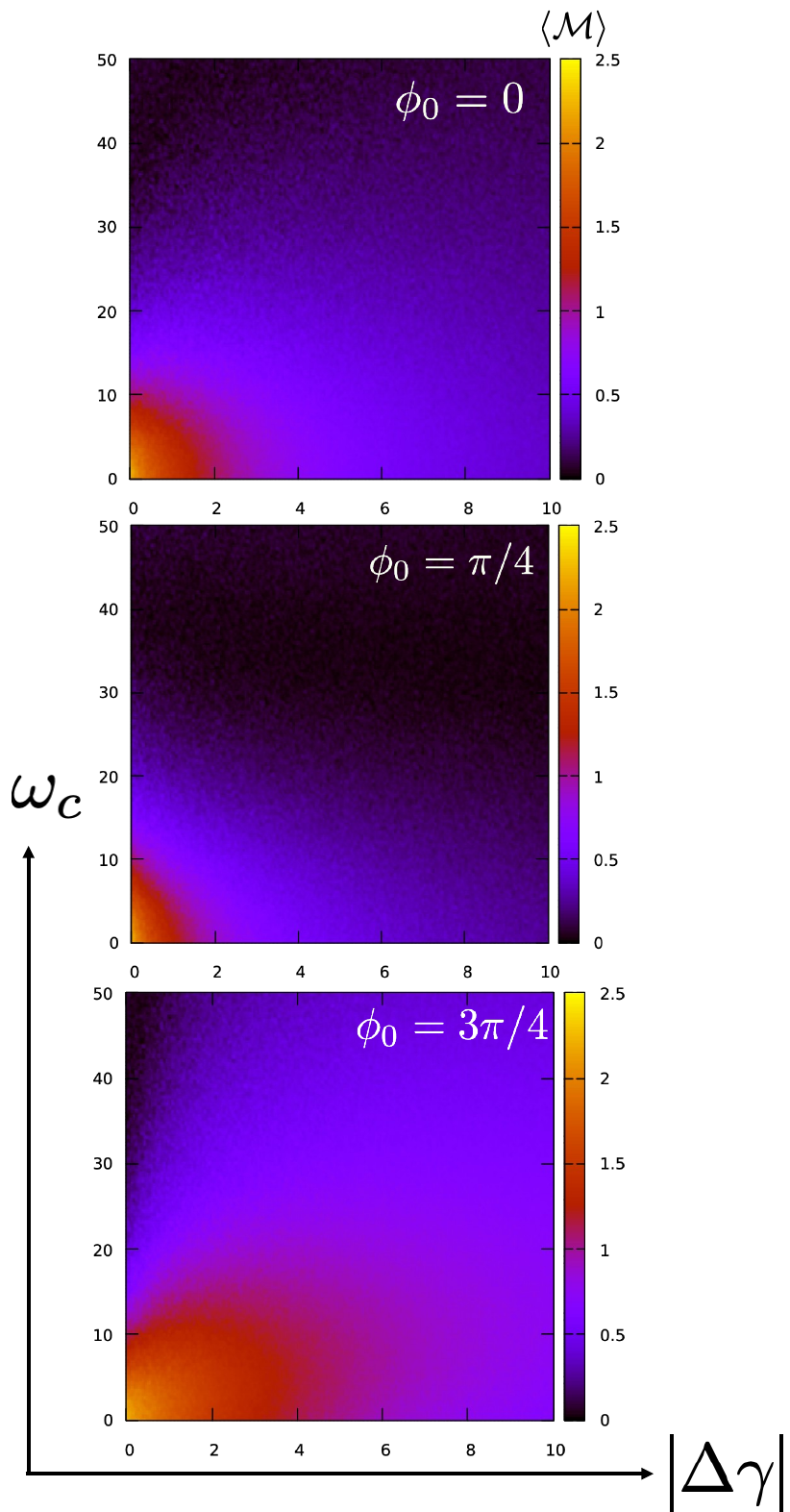}
    \caption{Parametric plots showing \emph{magnetic} and \emph{non-magnetic} phases, with $|\Delta\gamma|$ along $x-$axis, $\omega_c$ along $y-$axis and $\langle \mathcal{M} \rangle$ as color, for $\phi_0=0,\pi/4,3\pi/4$ respectively, and $|q|=1$. The rest of the parameters are kept as per the list mentioned earlier.}
\label{magmom}    
\end{figure}

Microscopic gyration can be manifested through orbital magnetic moment, when the gyrating particle is charged by an amount $q$ and subjected to an external, uniform magnetic field $\bf B$. Due to the gyration and also due to a cyclotron motion on the plane, here the charged ellipsoidal particle can form a current loop under $\bf B$ being applied perpendicular to the plane, thus generating a small, fluctuating orbital magnetic moment: $\mathcal{M} = \frac{|q|}{2} h $, directed normal to the plane, which can then be averaged to obtain: 

\begin{eqnarray}
  \langle \mathcal{M} \rangle = \frac{|q|}{2}\langle h \rangle  
\label{MagMom}
\end{eqnarray}

According to our previous discussion, as $\langle h\rangle$ depends on the shape anisotropy and inertia of the particle and also on the spherical asymmetry of the trapping potential, $\langle\mathcal{M}\rangle$ will also depend on them. Moreover, it will also depend on the cyclotron frequency, $\omega_c=\frac{|q|B}{m}$, of the system. Here, we will demonstrate how $\langle\mathcal{M}\rangle$ depends on $|\Delta\gamma|$ and $\omega_c$. This will eventually establish the geometry of the so-called \enquote{magneto-gyrator} as one of the knobs to control  $\langle\mathcal{M}\rangle$ , which can even be experimentally realized in dissipative systems like dusty plasma \cite{melzer2021physics,wang2018structures,kretschmer2005force}. In this context, one may also note that recently an isotropic magneto-gyrator has already been studied in \cite{abdoli2022tunable,muhsin2025active}. 

We will consider a charged ellipsoid gyrating across the $x-y$ plane and trapped by an asymmetric potential in the midst of two orthogonal temperatures, now being subjected to a uniform magnetic field along the $z$-axis. The additional Lorentz force will then modify the translational Langevin dynamics given in Eq.[\ref{eom1}] as:

\begin{eqnarray}
    m(\partial_t v_i)=-\gamma_{ij}v_j-\partial_i U + m\omega_c({\bf{v}}\times{\bf{\hat{z}}})_i + \xi_i(t)
    \label{eom4}
\end{eqnarray}

Together with $|\Delta\gamma|$, the velocity components of the particle are now also coupled by the magnetic field. The coupling between them is not equal anymore, as it was in the previous sections. Two circulations of the charged ellipsoid will now be prevalent in the system simultaneously --- namely, cyclotron motion and microscopic gyration. The net circulation of the charged particle at NESS can then be quantified by the average orbital magnetic moment, $\langle \mathcal{M} \rangle$. These two circulations will have their respective contributions to $\langle \mathcal{M} \rangle$. Though, in general, the average magnetic moment of the particle is influenced by all the system parameters such as shape, orientation, inertia, asymmetry parameter of the trapping potential, temperature difference, external magnetic field, etc., here we will focus only on the shape anisotropy, external magnetic field and mean orientation. We explore the aspects by numerically solving the translational dynamics (given by Eq.[\ref{eom4}]) and the orientational dynamics (given by Eq.[\ref{eom2}]) of the ellipsoid simultaneously, to determine the influence of the external field, inherent anisotropic shape of the corresponding magneto-gyrator and the mean orientation on the emergence of \emph{magnetic} $(\langle \mathcal{M} \rangle \neq 0)$ and \emph{non-magnetic} $(\langle \mathcal{M} \rangle = 0)$ phases. The work towards a more elaborative and systematic study on this system is in progress.


The results obtained from the numerical simulation of Eq.[\ref{eom4}] have been shown in the parametric plots of $\mathcal{\langle M\rangle}$ in the $|\Delta\gamma|-\omega_c$ plane (see Fig.[\ref{magmom}]). As per Eq.[\ref{h}], for a spherical charged particle undergoing gyration in the midst of zero magnetic field, the corresponding steady-state, average magnetic moment \cite{muhsin2025active} can be exactly calculated as: $\langle \mathcal{M} \rangle_\text{s,0}=\frac{|q|}{2}\left[\frac{\gamma\alpha(T_y-T_x)}{k\gamma^2 + m\alpha^2}\right]$, which agrees with our numerical results (residing at the origin of our parametric plots). For $\alpha=0$, the charged particle (irrespective of its shape) cannot gyrate in a harmonic trap and it is not in contact with the two heat baths simultaneously. Hence, the thermal reservoirs remain decoupled and the two translational DoF separately evolve towards their own unique thermal equilibrium. Thus, in the absence of a NESS, thermal equilibrium solely can never generate a net magnetic moment within the regime of classical physics, as per the \emph{Bohr–van Leeuwen theorem}.

For pure gyration $(\omega_c=0)$ of an ellipsoid, a non-zero magnetic moment persists for all possible values of $|\Delta\gamma|$, as the non-equilibrium heat flux survives via a non-zero $\alpha(T_y-T_x)$. The sign of $\langle \mathcal{M} \rangle$ is decided by the sign of $\alpha(T_y-T_x)$, which dictates the clock-sense of gyration, and consequently, the sign of resulting average orbital magnetic moment  of the system. On the other hand, for a spherical particle ($|\Delta\gamma|=0$) and with $\omega_c\neq0$, the magnetic moment decreases with an increasing magnitude of field \cite{muhsin2025active} . This occurs because of a net decrease in the circulation of the particle due to cyclotron motion and gyration happening simultaneously, thus competing with each other. Having discussed the basic premises and extreme limits of shape and field, we can now discuss the effects of mean orientation $(\phi_0)$ of the ellipsoidal magneto-gyrator on its emergent magnetic behavior.  As we know, $\gamma_{xy}$  is zero for $\phi_0=0$, and reaches its maximum positive value at $\phi_0=\frac{\pi}{4}=0.78$, and drops to the negative peak value at $\phi_0=\frac{3\pi}{4}=2.36$, and this trend is repeated periodically with the mean orientation. We will explore the behavior of $\langle \mathcal{M}\rangle$  in the plane of $|\Delta\gamma|$ and $\omega_c$, for these three extreme values of $\phi_0$. Thus, the qualitative features of the anisotropic magneto-gyrator, as depicted in Fig.[\ref{magmom}], can be explained as follows:

(i) For $\phi_0=0$, along the $\omega_c=0$ line, magnetic moment decreases with shape anisotropy (but does not enter into a non-magnetic state) and is prominent for completely spherical and slightly aspherical cases only. Along the $|\Delta\gamma|=0$ line, magnetic moment decreases with field and enters into a non-magnetic state at higher fields. However, with shape anisotropy and magnetic field both being non-zero, the magnetic moment persists even at higher anisotropy, but at lower field values.

(ii) For $\phi_0=0.78$, the fluctuations introduced in the system due to $|\Delta\gamma|$ will inhibit the gyration, resulting in a lower $\langle h\rangle$ (see Fig.[\ref{phi0}]), besides the inhibition of cyclotron motion at higher fields. Almost the entire parameter space will now be dominated by the non-magnetic state, except at low field and low anisotropy.

(iii) The profile of $\langle\mathcal{M}\rangle$ obtained for $\phi_0=2.36$  is considerably different. As shown in Fig.[\ref{phi0}], the specific angular momentum for gyration reaches its peak value due to a minimized velocity fluctuations. The inhibitory effect of high $\omega_c$ is subsided by the shape anisotropy here. The non-magnetic state has almost been eradicated and survives only for the spherical limit and around higher field values. However, the magnetic state survives even at higher field values, when the shape anisotropy is increased. In this case, the region of high $\langle \mathcal{M}\rangle $ is quite large in comparison to the other two cases. 

Further exploration of the various nuances of anisotropic magneto-gyrator can be done, using Eq.[\ref{eom2},\ref{eom4}] as the dynamical model. However, it is beyond the scope of the current paper.



\section{Concluding Remarks}

To conclude, we will summarize our results here. We have considered a single ellipsoid in 2D, with finite inertia, trapped in an asymmetric potential in a dissipative medium. The translational dynamics of the center-of-mass of the ellipsoid is governed by Langevin dynamics, where the velocity components of the center-of-mass are coupled by the tensorial friction while the position components are coupled via the asymmetry of the trapping potential. The orientational dynamics of the ellipsoid is governed by a restoring torque, which can be experimentally achieved by a precise opto-mechanical control. The ellipsoid is also simultaneously connected to two orthogonal heat baths, and therefore, it is driven to a NESS due to the resulting heat flux between the two heat baths. This NESS is manifested in the system as microscopic gyration, which essentially vanishes with either a null trap asymmetry or temperature difference. Here, with numerical simulation as well as analytic calculations, we have shown that the specific angular momentum of the gyrating ellipsoid has explicit dependence on the shape and orientation of the ellipsoid, which can be qualitatively different for high and low inertial regimes. The interplay of asymmetry of the trap, the inherent shape anisotropy, along with the inertia of the ellipsoid, has been extensively studied in the context of microscopic gyration. Various methods have been proposed to tune the relevant system parameters in order to obtain a better mean-to-fluctuation ratio, thus improving the gyratory response of the system. The analytical results are obtained from the steady-state covariance matrix, in the limit of small anisotropy in shape. Within the stipulated range of relevant parameters decided by the approximations, the numerical results related to the fully anisotropic system match well with the analytical expressions. With the introduction of inertia along with anisotropic shape in the dynamics of a gyrator, the inferences in this paper bear qualitative as well as quantitative differences with the conventional gyration obtained for overdamped as well as underdamped cases using a spherical particle \cite{bae2021inertial,muhsin2025active,abdoli2025enhanced,mancois2018two,dotsenko2013two,filliger2007brownian}.

Several qualitative as well as quantitative differences also exist between the gyration of an \enquote{overdamped} ellipsoid (as discussed in \cite{dutta2026microscopic}) and that of the \enquote{underdamped}/inertial ellipsoid discussed in this paper. We end by providing a brief comparative study between the two systems, on the basis of the common system parameters:

(i) In the overdamped case, a finite mobility difference helped in increasing the gyratory response, along with a simultaneous increase in the fluctuations. However, in the underdamped case, a decrease in the response due to the anisotropy of the friction is observed, which can be tuned by varying inertia of the ellipsoid. Also, the signal-to-noise ratio decreases monotonically with the anisotropy in friction in the underdamped case.

(ii) In the overdamped case, anisotropy of the potential is not required for steady-state gyration --- the role is substituted by the bi-axial shape itself. However, the anisotropic potential is necessary in the underdamped case, and the gyration shows a prominent dependence at higher potential anisotropy, which qualitatively alters with inertia.

(iii) The restoring torque stabilizes the gyration in the overdamped case, while the gyration becomes dynamically susceptible to this torque in the underdamped case --- the non-monotonic variation of mean specific angular momentum of the gyrating ellipsoid with the torque-strength being a confirmation to this.

(iv) Finally comes the mean orientation $(\phi_0)$ of the ellipsoid. In the overdamped case, the gyration is absent at $\phi_0=0$ and is maximized at $\phi_0=\frac{\pi}{4}$. Also, the periodic variation with $\phi_0$ changes the clock-sense of gyration. In the underdamped case, however, gyration is present even at $\phi_0=0$ and is maximized at $\phi_0=\frac{3\pi}{4}$, with no alteration in the clock-sense whatsoever.

\section*{Acknowledgments}

S.D. acknowledges University Grants Commission (UGC), India and University of Calcutta for financial assistance. A.S. acknowledges support from Anusandhan National Research Foundation (ANRF), DST (India), under the Advanced Research Grant (ARG) scheme (No.- DST(IN), ANRF/ARG/2025/009343/PS). A.S. also acknowledges support from CY Initiative of Excellence (Grant: “Investissements dAvenir” ANR-16-IDEX-0008) under CY Advanced Studies (CYAS), where the present work was partially developed. 

---------------------------------------------------------------

\section*{Appendix-I : Body-to-lab frame transformation of the equations of motion}

In the body frame (see Fig.[\ref{ellipsoid}]), at a temperature $T$, the underdamped Langevin dynamics corresponding to the translation of an ellipsoid can be written as:

\begin{eqnarray}
    \nonumber
    \dot{v}_X&=&-\frac{\gamma_\parallel}{m} v_X + \sqrt{\frac{2\gamma_\parallel T}{m^2}} \eta_X(t)\\
    \dot{v}_Y&=&-\frac{\gamma_\perp}{m} v_Y + \sqrt{\frac{2\gamma_\perp T}{m^2}} \eta_Y(t)
    \label{append1}
\end{eqnarray}

A small change in the velocity components in lab frame $(\delta v_x, \delta v_y)$ can be expressed in terms of the components in body frame $(\delta v_X, \delta v_Y)$ as: 

\begin{eqnarray}
    \nonumber
    \delta v_x&=& \delta v_X \cos\phi - \delta v_Y \sin\phi\\ 
    \delta v_y&=& \delta v_X \sin\phi + \delta v_Y \cos\phi
    \label{append2}
\end{eqnarray}

Dividing both sides of Eq.[\ref{append2}] by $\delta t \rightarrow 0$ will yield the acceleration components of the center-of-mass of the ellipsoid in the lab frame in terms of those in the body frame. One can now reuse Eq.[\ref{append1}] to obtain:

\begin{eqnarray}
    \nonumber
    \dot{v}_x&=&\left(-\frac{\gamma_\parallel}{m}v_X \cos\phi + \frac{\gamma_\perp}{m} v_Y \sin\phi\right) + \xi_x(t)\\
     \dot{v}_y&=&\left(-\frac{\gamma_\parallel}{m}v_X \sin\phi - \frac{\gamma_\perp}{m} v_Y \cos\phi\right) + \xi_y(t)
     \label{append3}
\end{eqnarray}

where, $\boldsymbol{\xi}(t)$ will be related to $\boldsymbol{\eta}(t)$ via the diffusivity tensor defined in Eq.[\ref{diff}]. But, Eq.[\ref{append3}] still contains the body velocities. One can get rid of them by using the inverse transformation of Eq.[\ref{append2}]. By doing so, one gets the elements of the friction tensor given by Eq.[\ref{gamm}].

It must be noted that the ellipsoid has a unique temperature $T$ in the body frame across all DoF, which is then experimentally modified in the lab frame such that $T_x \neq T_y$. Also, the confining potential $U(x,y)$ can be solely applied in the lab frame \cite{montana2023inertial}. After the inclusion of all these features, one can readily obtain Eq.[\ref{eom1}].

\section*{Appendix-II : Stochastic Energetics of the gyrator - the heat flux}

\subsection{Average heat flux --- Numerical results}

Here, we numerically obtain the average rate of the heat transfer to emphasize the fact that it indeed relates to the gyration. We will show how the rate of heat transfer depends on the anisotropies introduced in the system via an external potential and the geometry of the particle. The objective here is also to extend the previous studies on spherical gyrators \cite{bae2021inertial} by incorporating the effects of shape on the energetics of gyration and to provide additional insights on how the trap and shape asymmetries \enquote{compete} with one another, regarding the heat flux.  


The average rate of heat transaction (or, the heat flux) associated with the $i$-th DoF (with, $i\in\{x,y\}$) can be defined as \cite{sekimoto2010stochastic}: $\langle\dot{Q}_i\rangle=\langle\mathcal{F}_i\circ v_i\rangle$, where $\mathcal{F}_i$ is the total force generated from the reservoirs in which the ellipsoid resides, $v_i$ is given by Eq.[\ref{eom1}], and $\circ$ denotes the Stratonovich product of two stochastic quantities. From Eq.[\ref{eom1}], it is evident that the total force generated from the bath (that is, anisotropic viscous drag and the thermal noise) is exactly balanced by the inertial term and the anisotropic potential. Hence, the bath-generated force can be expressed as: $\mathcal{F}_i=m(\partial_t v_i)+\partial_i U(x,y)$ \cite{sekimoto2010stochastic}.

The non-zero heat flux, simultaneously subjected to $\alpha\neq0$ and $T_x \neq T_y$, is an indicator of the NESS prevalent in the system. We note that $\langle \dot{Q}_x + \dot{Q}_y \rangle=0$, which implies that no net power can be extracted from the system at the steady-state, that is, the heat absorbed from the hot bath is exactly dissipated to the cold bath via the gyrating ellipsoid. Net power can only be obtained from microscopic gyrators in the presence of non-reciprocal forces or time-dependent drives, as proposed in \cite{movilla2021energy,miangolarra2022thermodynamic}. Nonetheless, for $T_x>T_y$, the heat reservoir along $x$ acts as the \emph{hot} bath, while the one along $y$ acts as the \emph{cold} bath. This is also confirmed by the usual sign convention, whereby $\langle \dot{Q}_x \rangle >0$ and $\langle \dot{Q}_y \rangle <0$ --- the gyrator takes heat from the bath along $x$ and dissipates it to the bath along $y$. The exchanged heat between the two orthogonal baths vanishes for $\alpha=0$. 

\begin{figure}[htp]
    \centering
    \includegraphics[width=6cm]{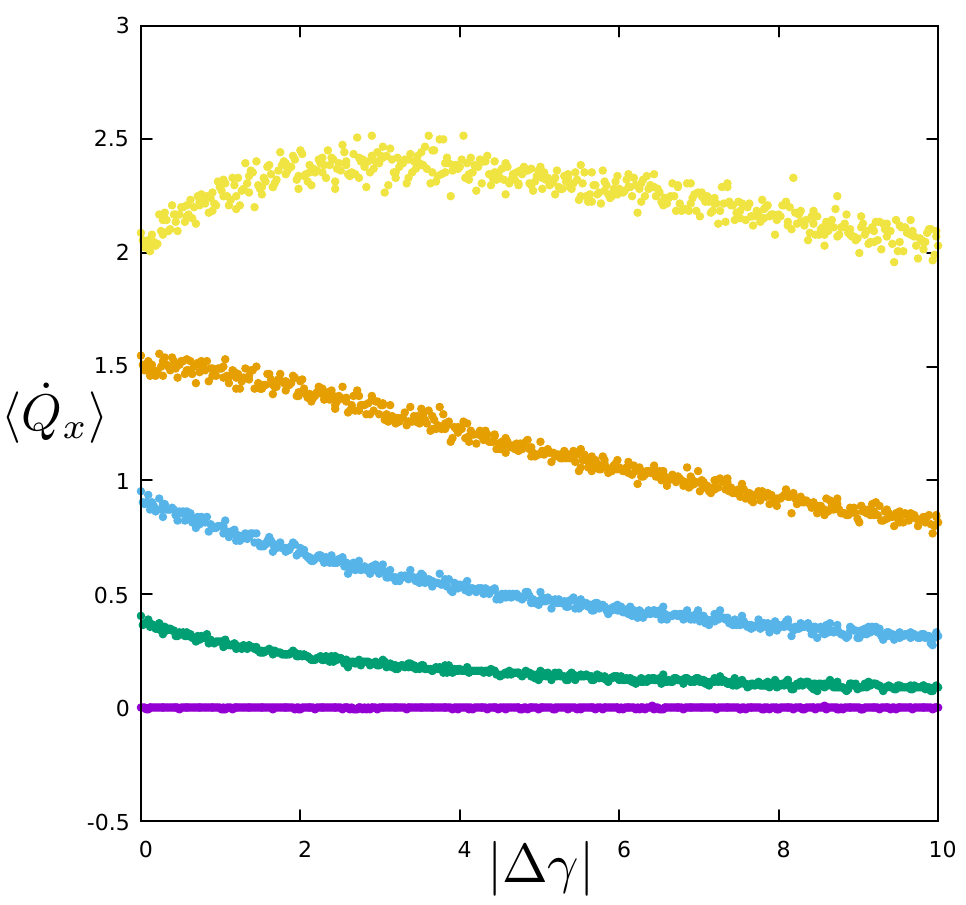}
    \caption{Plot of $\langle \dot{Q}_x \rangle(>0)$ versus $|\Delta\gamma|$, for various $\alpha$ : the purple, green, blue, orange and yellow points are for $\alpha=0,0.3,0.5,0.7,0.9$ respectively. Here, $T_x=10$ and $T_y=1$.}
\label{heatvsdeltagamma}    
\end{figure}

Fig.[\ref{heatvsdeltagamma}] shows the variation of $\langle \dot{Q}_x \rangle$ with $|\Delta\gamma|$, for $T_x > T_y$. In the spherical limit $(|\Delta\gamma|=0)$, the values of heat flux (for different $\alpha$) are consistent with the analytical expression in \cite{bae2021inertial}. The average heat flux becomes zero for $\alpha=0$, whereby the gyration ceases to exist. For $\alpha$ being non-zero, the heat flux also becomes non-zero and the gyration starts. We emphasize that when $\alpha$ is non-zero, gyration survives in the spherical limit ($|\Delta\gamma|=0$) as well. For smaller, non-zero values of $\alpha$, the heat flux monotonically decreases with an increase in $|\Delta\gamma|$. This occurs because $\alpha$ being small, the heat flux between the baths are reduced (causing $\langle h\rangle$ to be small) and moreover, as $|\Delta\gamma|$ couples $v_x$ and $v_y$, their fluctuations get coupled as well, which eventually reduces the mean heat flux with an increasing $|\Delta\gamma|$. As $\alpha$ approaches $k$, the variation changes qualitatively. The heat flux now varies non-monotonically with $|\Delta\gamma|$, and decreases only after reaching a peak value at some non-zero $|\Delta\gamma|=|\Delta\gamma|^*$ (when $\alpha$ becomes as large as $0.9$, with $k=1$), starting  from $|\Delta\gamma|=0$. For $\alpha\rightarrow k$, both $\langle xy \rangle$ and $\langle h \rangle$ become large. Here, the gyration is largely facilitated by $\alpha$ but inhibited by $|\Delta\gamma|$. Thus, $\langle \dot{Q}_x \rangle$ is optimized in the midst of these two competing factors, and we obtain its non-monotonic variation with the shape anisotropy of the ellipsoid.

\begin{figure}
    \centering
    \includegraphics[width=6cm]{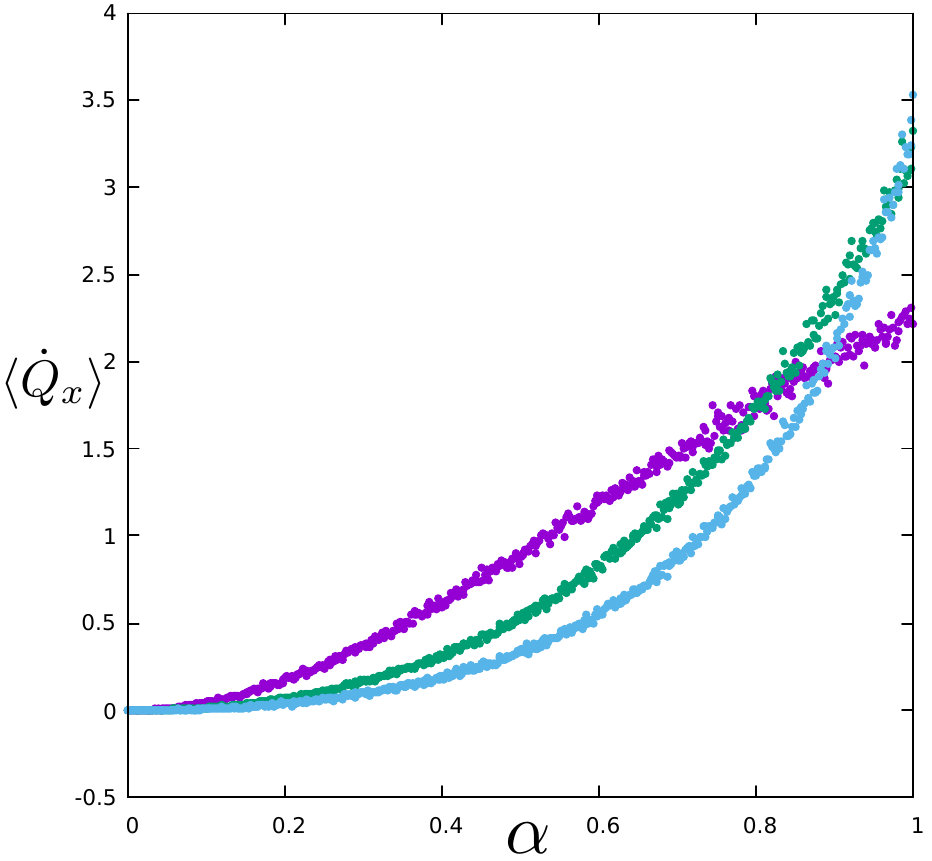}
    \caption{Plot of $\langle \dot{Q}_x \rangle(>0)$ versus $\alpha$, for various $|\Delta\gamma|$ : the purple, green and blue points are for $|\Delta\gamma|=0,4,9$ respectively. Here, $T_x=10$ and $T_y=1$.}
\label{heatvsalpha}    
\end{figure}

Fig.[\ref{heatvsalpha}] shows the variation of $\langle \dot{Q}_x \rangle$ with $\alpha$, for $T_x > T_y$. $\langle \dot{Q}_x\rangle$ shows qualitative differences between the spherical and ellipsoidal cases, as evident from the nature of its variation with $\alpha$. For an initial range of close-to-zero values of $\alpha$, all the three cases merge to a null value of the heat flux, due to a small or negligible gyratory response. As $\alpha$ is increased towards unity, for the spherical case $(|\Delta\gamma|=0)$, the heat flux varies as $\sim\frac{\alpha^2}{1+c\alpha^2}$,  with a constant $c$. When $|\Delta\gamma|$ is switched to non-zero values, the dependence becomes $\sim\frac{\alpha^2}{1+c_1\alpha+c_2\alpha^2}$, where $c_1$ and $c_2$ are constants. This leads to a slower variation of the flux at smaller $\alpha$ and a rapid variation with larger $\alpha$. These functional forms of $\langle \dot{Q}_x\rangle$ w.r.t. $\alpha$ will also be supported by the approximated, analytical results in the next section. One may also note the fact that the values of flux for $|\Delta\gamma|=9$ are almost always smaller than those of $|\Delta\gamma|=4$, indicating the inhibitory effects of shape on the gyration (as discussed earlier). Interestingly, for $\alpha\rightarrow k(=1)$, $\langle \dot{Q}_x\rangle$ becomes larger as compared to the spherical case.

\subsection{Average heat flux --- Approximated analytical result}


From Eq.[\ref{eom3}], the bath-generated force can be explicitly written (for $x$, say) as: $\mathcal{F}_x=m {(\partial_tv_x)} + (kx+\alpha y)$, obeying Sekimoto's prescription \cite{sekimoto2010stochastic}. In the steady-state, the heat fluxes can now be obtained (using Eq.[\ref{covariance}]) as:

\begin{eqnarray}
    \langle \dot{Q}_x \rangle = \frac{\gamma\alpha^2 (T_x-T_y)}{2[\gamma^2(k+\alpha|\epsilon|)+m\alpha^2]} = - \langle \dot{Q}_y \rangle
    \label{heat}
\end{eqnarray}

where, $|\epsilon|=0$ yields the spherical case \cite{bae2021inertial}. For the slightly-anisotropic limit, however, the qualitative dependence of the heat flux on the potential anisotropy $(\alpha)$ changes due to the introduction of a new term $(\sim \alpha |\epsilon|)$ in the denominator of Eq.[\ref{heat}]. This expression can also serve as a measure of thermal \enquote{conductivity} of the gyrator.

As discussed earlier, the heat flux vanishes for either $\alpha=0$, or, $T_x = T_y$ --- the causal factors for gyration to occur. We also note that $\langle \dot{Q}_x + \dot{Q}_y \rangle=0$, which implies that the heat absorbed from the hot bath is exactly dissipated to the cold bath. For $T_x>T_y$, the heat reservoir along $x$ acts as the hot bath, while the one along $y$ acts as the cold bath. This is now analytically confirmed by the usual sign convention, whereby $\langle \dot{Q}_x \rangle >0$ and $\langle \dot{Q}_y \rangle <0$. The heat fluxes for the complete anisotropic system have been numerically studied in Figs.[\ref{heatvsdeltagamma},\ref{heatvsalpha}].

\bibliographystyle{unsrt}
\bibliography{sample.bib}

\end{document}